\documentclass[preprint,12pt]{article}
\usepackage[a4paper, total={6.5in, 8.5in}]{geometry}

\usepackage[normalem]{ulem}

\usepackage{amsmath, amssymb, amsthm, bbm}
\usepackage{enumitem}
\usepackage{tikz}
\usepackage{graphicx}
\usepackage{xcolor}
\usepackage{dsfont}
\usepackage{booktabs} 
\usepackage[UKenglish]{isodate}
\usepackage{graphicx}
\usepackage[font=small,labelfont=it]{caption}
\usepackage{subcaption}
\usepackage{hyperref}
\usepackage{enumitem}
\usepackage{comment}
\usepackage{multirow}
\usepackage{fancyvrb}
\usepackage{footmisc}
\usepackage{tikz-cd}
\usepackage{textcomp}
\usepackage{textcomp}
\usepackage{eurosym}

\renewcommand{\phi}{\varphi}

\newtheorem*{theorem*}{Theorem}
\newtheorem{theorem}{Theorem}[section]

\newtheorem{remark}[theorem]{Remark}

\NewDocumentCommand{\luca}{mo}{
    \IfValueF{#2}{
                        {{\scriptsize
                            \textcolor{blue}{ 
                            \textbf{L:}
                            \textit{{#1}}
                            }
                        }}
        }
    \IfValueT{#2}{
                        \marginnote{{\scriptsize
                            \textcolor{blue}{ 
                            \textbf{L:}
                            \textit{{#1}}
                            }
                        }}
        }
                    }
\NewDocumentCommand{\giulia}{mo}{
    \IfValueF{#2}{
                        {{\scriptsize
                            \textcolor{black}{ 
                            \textbf{GL:}
                            \textit{{#1}}
                            }
                        }}
        }
    \IfValueT{#2}{
                        \marginnote{{\scriptsize
                            \textcolor{black}{ 
                            \textbf{GL:}
                            \textit{{#1}}
                            }
                        }}
        }
}
\NewDocumentCommand{\anastasis}{mo}{
    \IfValueF{#2}{
                        {{\scriptsize
                            \textcolor{jade}{ 
                            \textbf{A:}
                            \textit{{#1}}
                            }
                        }}
        }
    \IfValueT{#2}{
                        \marginnote{{\scriptsize
                            \textcolor{jade}{ 
                            \textbf{A:}
                            \textit{{#1}}
                            }
                        }}
        }
                    }

\NewDocumentCommand{\lukas}{mo}{
	\IfValueF{#2}{
		{{\scriptsize
				\textcolor{magenta}{ 
					\textbf{Lukas:}
					\textit{{#1}}
				}
		}}
	}
	\IfValueT{#2}{
		\marginnote{{\scriptsize
				\textcolor{magenta}{ 
					\textbf{Lukas:}
					\textit{{#1}}
				}
		}}
	}
}

\title{Understanding the householder solar panel consumer: a Markovian model and its societal implications}
\author{ Marta~Leocata\thanks{Luiss University, Viale Romania 32, 00197, Roma, Italy. Email address: mleocata@luiss.it}
\and Giulia~Livieri\thanks{The London School of Economics and Political Science, London, United Kingdom. Email address: g.livieri@lse.ac.uk}
\and Silvia~Morlacchi\thanks{Dipartimento di Matematica, Università di Pisa, Largo Pontecorvo 5, 56127, Pisa, Italy. Email address: silvia.morlacchi@dm.unipi.it}
\and Fausto~Corvino\thanks{Hoover Chair in Economic and Social Ethics \& Institut supérieur de philosophie, Université catholique de Louvain, Place Montesquieu 3, 1348, Louvain-la-Neuve, Belgium. Email address: fausto.corvino@uclouvain.be}
\and Franco~Flandoli\thanks{Scuola Normale Superiore, Pisa, Italy. Email address: franco.flandoli@sns.it}
\and Alberto~Pirni\thanks{Scuola Superiore Sant’Anna, Pisa, Italy. Email address: alberto.pirni@santannapisa.it}}

\date{\today}

\begin{document}
\maketitle
\begin{abstract}
\textcolor{black}{\noindent Household adoption of rooftop photovoltaic (PV) systems is central to the green energy transition, yet diffusion depends on social influence and behavioral biases, as well as payback economics. This study develops a parsimonious Markovian model in which households move sequentially from being unengaged (``Carbon") to informed, to planning, and finally to adoption (``Green"). Transition rates are micro-founded by two mechanisms: (i) social contagion/communication, proxied by the current share of adopters, and (ii) economic profitability, proxied by payback time computed from a Net Present Value framework. Novel to this diffusion setting, bounded rationality is introduced via hyperbolic discounting, creating a procrastination loop that delays adoption even when PV is economically attractive in a long-run perspective. Calibrated on
the Italian residential PV diffusion path (2006–2020) and assessed in national and regional applications, the model reproduces observed trajectories and enables forward-looking scenario analysis (2020–2026). Results show that policies yielding similar payback improvements can produce different outcomes once present bias is accounted for and that behaviorally informed intervention are stronger. The findings contribute
a micro-to-macro bridge between behavioral economics and technology diffusion modeling and imply that effective policy portfolios (and PV business models) should complement incentives with commitment devices and social-norm peer strategies to accelerate PV uptake and its spillover emissions benefits.}
\end{abstract}
{\small
{\bf Keywords:} green energy transition, solar photovoltaic, individual based modelling, ethics and mathematics, procrastination, Markovian model.\\
{\bf 2020 Mathematics Subject Classification:} 60K35, 91A16, 60J27. \\
}

\newpage
\section{Introduction}\label{sec:Introduction}
\subsection{Motivation of the work and modelling framework.}\label{subsec:MotivationAndModelling}
\begin{figure}
    \centering
        \includegraphics[scale=0.45]{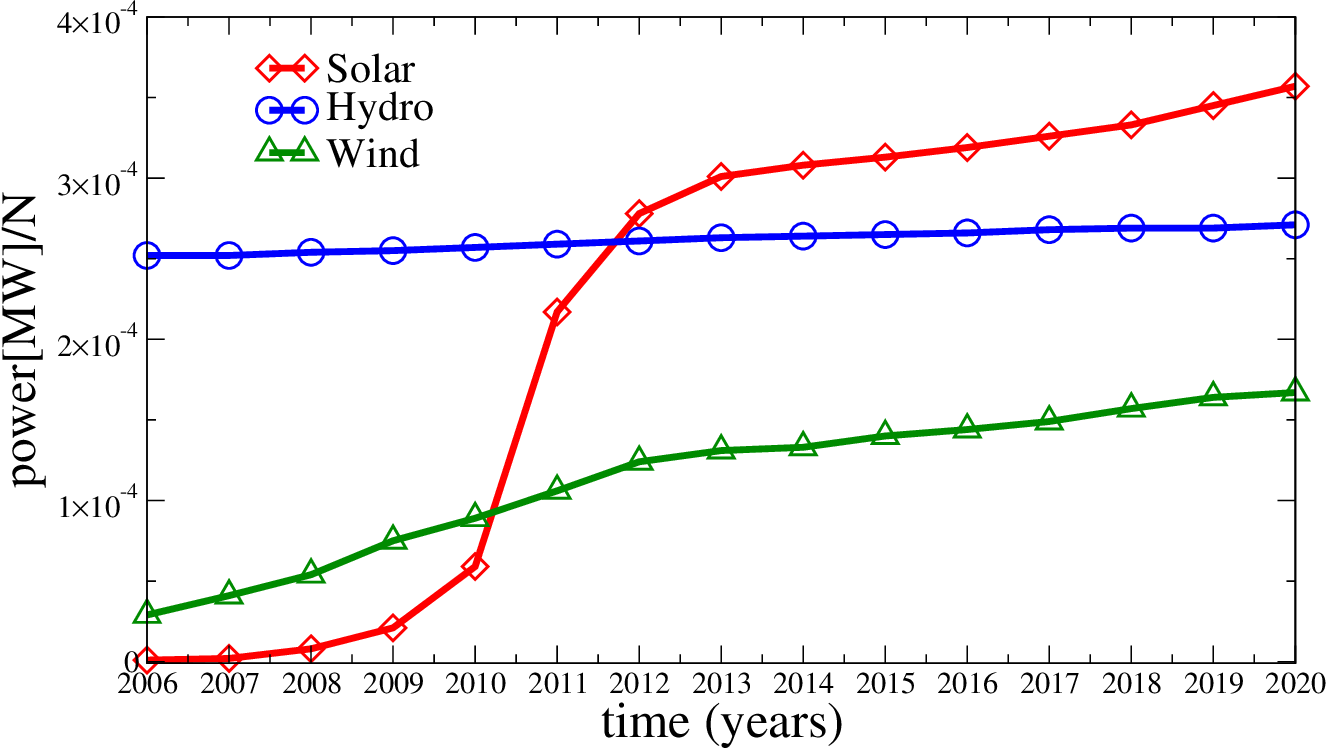}
        \caption{Evolution of Italy's renewable electricity production by sources: ``Solar" (\textit{red line}), ``Hydro" (\textit{blue line}), ``Wind" (\textit{green line}) over 2006--2020. \textit{Data Source\,:\,} Terna Spa (\url{https://www.terna.it}).}
        \label{fig:RenewableElectricity}
\end{figure}

\begin{figure}
    \centering
        \includegraphics[scale=0.45]{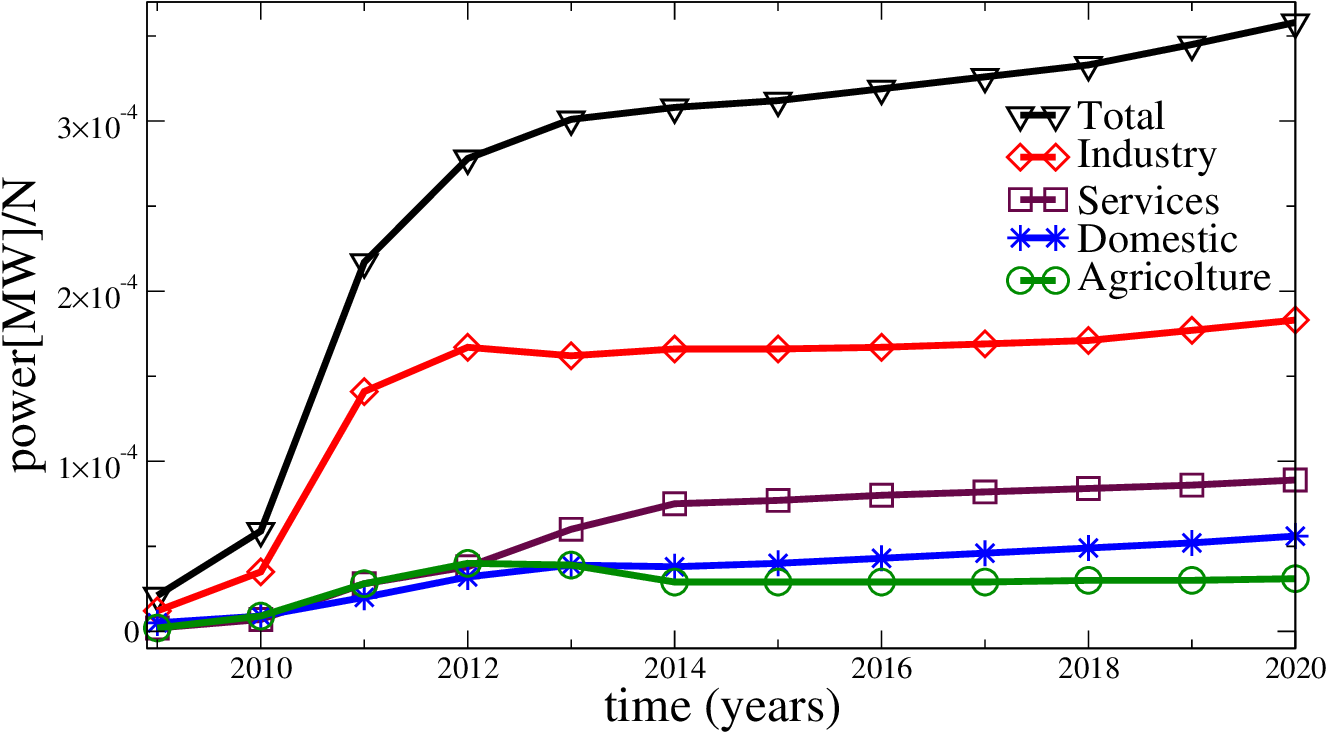}
        \caption{Evolution of Italy's electricity production, source ``Solar", by categories: ``Agriculture" (\textit{blue line}), ``Domestic" (\textit{black line}), ``Services" (\textit{red line}), ``Industry" (\textit{green line}) over 2006--2020. \textit{Data Source\,:\,} GSE (\url{https://www.gse.it/dati-e-scenari/statistiche}).}
        \label{fig:RenewableElectricity2}
\end{figure}

\textcolor{black}{One of the top challenges of this century is reducing greenhouse gas emissions and preventing dangerous interferences with the climate system. In 2015, governments committed to drastically reducing their emissions under the Paris Agreement but faced the challenge of turning pledges into practical policies.   Against the obstacles faced by several \textit{carbon pricing} policy proposals (e.g., \cite{Carat2015}), some jurisdictions turned to subsidies for renewable energy as an alternative to  ``first-best" policies. In particular, generating electricity from solar photovoltaic systems has played an essential role in the transition towards an energy system based on renewable (\cite{IEA2019}). It contributes to meeting the climate change mitigation scenario in which global temperature rise is kept within 1.5 degrees Celsius (\cite{IPCC2018}).} \textcolor{black}{Solar energy is being utilized worldwide, with over 126 countries implementing strong policies and regulatory frameworks to promote its growth
(\cite{IEA}). The increase in solar energy generation has been remarkable, rising from just 31 terawatt-hours (TWh) in 2010 to over 1,000 TWh in 2021 -- a growth of more than 30 times in just a decade (\cite{IRENA}). Despite this impressive progress, it is widely acknowledged that our current usage of solar energy is still significantly below its vast potential, highlighting the urgent need for targeted initiatives to enhance its adoption (\cite{shakeel2023solar}).\\ 
\indent Recent research has emerged from this awareness, with experts systematically exploring the critical factors influencing the adoption of solar photovoltaic systems at the household level (see, e.g., the two very recent literature reviews \cite{alipour2021residential, shakeel2023solar}). Notably, while Italy is a significant player in solar energy generation and installed capacity -- alongside countries like France, Germany, Greece, Spain, and the U.K. (e.g., \cite{dusonchet2015comparative}) -- it has not been sufficiently examined in the existing literature (e.g., \cite{shakeel2023solar}, Table 4, and \cite{bianco2021analysis}). For instance, in \cite{kriechbaum2018looking}, authors analyze the photovoltaic panels adoption in Germany and Spain. It is claimed that a hype and subsequent phase of disappointment, both associated with a level of expectation specific for the country, occurred. However, the perspective in the previous study is slightly different from what is proposed in this work: \cite{kriechbaum2018looking} focuses on the spreading of collective expectations by analyzing newspapers attention, while in our model it is assumed that the adoption decision is influenced by the network communication of each person or family and the so-called payback period of the investment. In particular, the choices made by one individual influence the likelihood that others will make similar choices, with a contagion effect. On the other hand, the imitation phenomenon has been taken into account in, e.g.,  \cite{dong2017forecasting} and \cite{guidolin2010cross}. The former study focuses on the diffusion of the residential solar photovoltaic system in California by employing a time-series forecasting model, a threshold heterogeneity diffusion model, a Bass diffusion model, and National Renewable Energy Laboratory’s Solar model. The latter, instead, focuses on the diffusion of solar photovoltaic systems in many countries, Italy being one of them. However, it does not take into account the recent regional data splitting, which allows our analysis to focus also on this aspect of the diffusion dynamics. Moreover, its analysis concerns the installed photovoltaic solar power, and not the number of installed systems; while in our study, in order to focus on the residential sector, we focus on the number of installed units. Anyway, the bounded-rationality of agents is not taken into account; see below.\\
\indent The fact that Italy has not been sufficiently examined in the existing literature presents a compelling opportunity for further investigation, particularly since Italy is an interesting case study: It is a large electricity market with optimal climate conditions and, more importantly, has rapidly become one of the leading EU countries in terms of photovoltaic installed capacity, even starting from a point where solar energy contribution was virtually nonexistent (see Figure \ref{fig:RenewableElectricity}). In addition, the aforementioned studies operate under the assumption of perfect rationality among householders. However, it is known that the actual decision-making behaviors of individuals or families often diverge from the theoretically optimal behavior predicted by perfect rationality, especially in the renewable energy sector (e.g., \cite{li2025dynamic, masini2013investment, cao2023distributionally, he2024optimized}).\\
\noindent We reference several agent-based model studies in the electricity sector that account for the bounded rationality of the involved agents (see \cite{barazza2020impact}, Table 1): \cite{kwakkel2014exploratory, kraan2019jumping, chappin2017simulating, kraan2018investment}. However, these studies primarily focus on electricity generators (\cite{kwakkel2014exploratory, chappin2017simulating}) or investors in existing and new electricity sectors (\cite{kraan2019jumping, kraan2018investment}), situated within two interconnected electricity markets in typical European countries (\cite{chappin2017simulating}), the Netherlands (\cite{kwakkel2014exploratory}), and in liberalized European markets (\cite{kraan2019jumping, kraan2018investment}). Additionally, the overwhelming majority of these papers are \emph{a-theoretical}, meaning that they lack a solid theoretical foundation for their research design; they have predominantly relied on existing literature to develop frameworks. Specifically, authors in \cite{shakeel2023solar} have highlighted a clear need to broaden prevailing cross-disciplinary approaches by incorporating a wider variety of theoretical concepts to improve the construction of empirical and conceptual studies.\\
\indent Based on the analyzed contributions, there appears to be a significant opportunity to address the gap in understanding the Italian context by proposing a grounded theory model that also considers the impact of householders' bounded rational behaviors on the adoption of solar photovoltaic systems. In particular, we will focus on \textit{procrastination} as a behavioral bias that complicates the promotion of energy efficiency renovations. This bias has primarily been studied in the fields of health and financial savings but has received only limited attention in energy saving research (e.g., \cite{lillemo2014measuring, lades2021maybe}). In this regards, we mention the very recent paper \cite{mogensen2024stop}, where authors conducted a survey of 609 Danish homeowners in the final stage of an energy efficiency renovation decision-making processes, and through a comparative analysis of adopters and non-adopters, they find that procrastination is a common reason for non-adoption.}
\textcolor{black}{In detail}, this paper aims to understand how Italy's \textit{public sphere} has behaved on the theme of green energy transition (GET, henceforth), where for the public sphere, we mean \textit{individual people} or \textit{families}, which we call \textit{agents} hereafter. Notably, the term agents is used in many areas of research to exclusively indicate individuals whose actions (also named controls) are the result of an optimization process. On the contrary, in \cite{FCLLMP2022ArXiv}, the authors show that a Markovian-type framework where agents do not make rational decisions based on optimization rules is more suitable to describe the public sphere's behaviour on the GET, where people occasionally question themselves about the GET problem; \textcolor{black}{see also the discussion below regarding the description of Figure \ref{fig:RenewableElectricity} and Figure \ref{fig:RenewableElectricity2}.}    \textcolor{black}{In addition, we mention} the extensive literature on opinion dynamics in which the evolution of opinions in society is modelled through Markov chains; see, e.g., \cite{GME2011JOSP,HL1974AOP,LNL1992PRA,SLST2017PSOCA}.\\
\noindent \textcolor{black}{In light of this discussion, we adopt a Markovian-type framework in the present paper. However, no matter how promising this approach may appear,  we are not pretending that it is sufficient ``as is" for chasing \textit{black swans}, i.e., unpredictable events beyond what is typically expected of a situation and have potentially severe consequences. Instead, the models presented in Section \ref{sec:MarkovianModels} analyze events whose broad dynamics are quite well understood. Furthermore, simulations in Section \ref{sec:numerical_experiments} contribute to putting in evidence the role of some key parameters. They can indicate how to devise external actions to obtain a specific behaviour of the society under consideration. Accounting for black swan in a Markovian-type framework is beyond the scope of the present paper, though an exciting venue for future research; a promising starting point could be the very recent working paper \cite{lee2024black}.}\\
\indent There is a growing acknowledgement in the literature and practice that, despite energy technology being available, and in many cases economically beneficial, other barriers prevent households' widespread adoption of new green technologies (see, e.g., \cite{LUTHRA2014RenAndSust}). In particular, \textcolor{black}{we focus our discussion mainly on the following two facts:} (1) the inclination of humans to \textit{mimic} the behaviour of other people; (2) the natural inclination of humans to \textit{procrastinate}. Among the GET examples, we focus, as said, on the case of photovoltaic systems (PVs, henceforth), the primary motivation being the availability of a \textit{relatively} significant sample of data; see Subsection \ref{subsec:datadescription}. The present article provides conceptual and empirical results to better understand agent behaviour in the solar photovoltaic market. In particular, the present paper aims to answer the following research question: What are the factors playing an essential role in the decision process for PVs?\\ 
\indent \textcolor{black}{Before turning to the presentation of our models, we display in Figure \ref{fig:RenewableElectricity} Italy's renewable electricity production by sources over 2006--2020, and in Figure \ref{fig:RenewableElectricity2} the ``Solar" time-series decomposed among the following four categories: ``Agriculture," ``Domestic," ``Services," and ``Industry," with the ``Domestic" one being our main focus (see the discussion above). Importantly, a subdivision into two periods, separated by 2012, characterizes ``Industry" and ``Domestic". Both time series show a relatively exponential solid increase in the period before 2012 and a weaker exponential increase in the period after 2012. In addition, “Industry” has a solid increase, roughly linear, around 2011, essentially absent in the ``Domestic" time-series; in 2011, Italy proposed a much more substantial Feed-in-Tariff subsides; see Appendix \ref{app:subsec:evolution_PVs} for a description of the evolution of Italy's solar photovoltaic market. In \cite{FCLLMP2022ArXiv}, the authors show that a game theory model explains the linear increase of 2011 observed for ``Industry". In contrast, a Markovian model easily fits the exponential increase periods. However, it is not natural to explain the solid linear increase of 2011, which, as said, is not present in the ``Domestic" time series. In other words, companies, around 2011, underwent a game; indeed, 2011 came after some years of moderate-size FiT subsidies.   In 2011,    Italy proposed a much more substantial FiT subsidy. Companies acted as in a game, whereas single people or families were not prepared; they reacted but not with the game's logic.    This fact is mainly because the planning ability of companies is superior to that of domestic ones. Whence, it is more natural to use a Markovian model in explaining ``Domestic" behaviour.} 

\newpage
\textcolor{black}{Notice that the decision to adopt photovoltaic (PV) technology has a long-term aspect to it, which entails upfront costs but ultimately pays for itself over time through the generation of electricity with minimal variable expenses, primarily linked to maintenance. We consider two Markovian models; in particular, the second model is a refinement of the first one.}  In both models, a variable $X_t^{i}$ characterizes the state of each agent \textit{i}, \textit{i}$\in \{1, \ldots, N\}$, being $N$ the number of agents. A value of $X_t^{i} = C$ stands for ``Carbon" and indicates that the individual has not decided on the theme of GET yet. She/he can be either agnostic to such a theme or prone to change her/his mind. For instance, she/he can be prone to mimicry, easily swayed by the behaviours of others in her/his social group, and attentive to social power and hierarchy. A value of $X_t^{i} = I$ stands for ``Informed" and characterizes an agent that is fully informed on the benefits of PVs or that has developed a certain level of sensitivity to climate change and environmental issues\footnote{Notice that, generally, having information or being environmentally motivated may either not coincide or be equivalent: An agent can be highly informed but do nothing or know very little but be highly motivated. However, we will leave the modelling of the previous situations for future research.}. Finally, $X_t^{i} = G$ stands for ``Green" and denotes the state of the individual that has installed the PVs.\\
\indent In the first model, we assume that each agent $i$ can only pass from the state ``Carbon" to ``Informed", and from the state ``Informed" to ``Green"; the precise mechanism under which the transition takes place is described in Section \ref{sec:MarkovianModels}. In the second model, more structure is added to the transition from the state ``Informed" to the state ``Green." This additional structure hinges on the concept of \textit{procrastination}, which means postponing into the future something that, from a subjective perspective, it would be rational to do earlier (\cite{A2010Book}). More precisely, we add the state $PL$ to the variable $X_t^{i}$ to capture such a behaviour. The value $X_t^{i}=PL$ stands for ``Planner" and the transition from ``Informed" to ``Planner" captures the following behaviour. \textcolor{black}{In general, not all individuals who have not adopted PV can be labeled as procrastinators. Various rational factors might explain the delay, such as financial constraints, difficulty in accessing credit, lack of information about available options, housing circumstances, risk
misperception, etc. Those who overcome these barriers and develop a clear preference for purchasing PV, i.e., have become ``Informed" in our language, may still find themselves in a procrastination loop due to the long-term nature of the decision. As the time to make the purchase nears, the perception of investment costs versus future energy benefits can fluctuate, leading to a temporary reversal of preferences, typically followed by regret, akin to what occurs in instances of procrastination in general. Once the agent gets out of the procrastination loop, she/he passes from ``Informed" to ``Planner" and then, as before, from ``Planner" to ``Green." Again, the precise mechanism under which the transition occurs is described in Section \ref{sec:MarkovianModels}.\\
\indent Finally, in both models,} we assume, as predominantly done in behavioural economics, that the final transition from ``Informed" to ``Green" or from ``Planner" to ``Green" happens in response to a cost-benefit ratio. We assume that each agent makes the \textit{final} choice of installing PVs based on benefits outweighing costs. More precisely, we assume that our agents first become an \textit{homo sustinens} agents (\cite{GAM2021EN}), and then (necessarily) \textit{homo economicus} agents before making the transition to the state ``Green"; see Table 3 in \cite{GAM2021EN} for a nice overview of these two types of agents.  \textcolor{black}{More precisely, our agents are \textit{boundedly rational homo economicus} agents; we will defer the discussion on this topic to the end of Section \ref{sec:MarkovianModels}.} Admittedly, one can construct a refinement of the second model in which some agents can pass from ``Informed" to ``Green," but we leave this extension for future research (we will return to this point later). We also mention two other possible refinements that we would like to consider in the future. First, we want to include the possibility of the transition from ``Informed" to ``Carbon" occurs. In other words, we desire to include the mechanisms that cause an agent's opinion to revert to the initial one. Another significant point would be the 
 \textcolor{black}{explicit} inclusion of bureaucratic obstacles to installations, such as slow installation and difficulty finding technical information. 
\subsection{Positioning in the existing literature.}\label{subsec:LiteratureAndContribution}
\indent In his article, Gifford (\cite{G2011AP}) proposes a framework to describe why humans are not taking action to prevent or ameliorate climate change. Energy inefficiency is a similarly complex and abstract problem to climate change. In particular, Gifford postulates the following seven ``dragons" of inaction with regard to climate change: (1) ``limited cognition," (2) ``ideologies," (3) ``dis-credence," (4) ``perceived risk," (5) ``sunk costs," (6) ``comparison with others," (7) ``limited behaviours". So far, different authors have tried to analyze or incorporate (some of) these dragons into mathematical models through the lenses of different approaches in order to fit PV data. Here, we mention the following works, which do not represent, however, a comprehensive list. (a) Survey-based analyses; see, e.g., \cite{CAM2021ERSS}. (b) Finite element methods to account for spatial heterogeneity; see, e.g., \cite{K2016AE}. (c) Variants of the popular Bass' model (\cite{B1969MS}); see \cite{DS2020EFSD} which state that the diffusion of solar photovoltaic systems in Brazil is highly influenced by the knowledge about such systems. (d) The agent-based \textcolor{black}{simulation} approach of, e.g., \cite{ZMCS2022SMPT}, \cite{PSM2015TFSS}, and \cite{PBOL2022AE}. The agent-based approach offers a framework to explicitly model the adoption decision process of the agent of a heterogeneous social system based on their individual preferences, behavioural rules, and interaction/communication within a social network. In particular, in the previous works, it is assumed that each agent decides to install a PVs at a certain time $t$ when his/her total utility at that time is greater than a certain threshold, usually calibrated on data. For instance, in the very nice work of \cite{PSM2015TFSS}, the total utility equals the sum of four weighted partial utilities accounting respectively for the payback period of the investment, the environmental benefit of investing in a PV system, the household's income, and the influence of communication with other agents. Therefore, these utilities concur \textit{at the same time} to determine whether or not an agent adopts a PV system.\\
\indent \textcolor{black}{Our modelling framework can also be considered an agent-based model, but, importantly, the name agent-based in our case is very close to the term ``interacting particle" commonly used in Physics, where the particles are all equal and subject -- more or less -- to the same rules of interaction among them and with the environment. However, here, the particles are individual people or families, which have more complexity than particles in Physics and have the possibility of making decisions. The main difference with respect to the agent-based simulation approach described in the previous paragraph is because we assume that the number of particles is huge, but the number of parameters is relatively small since the interaction and decision rules are the same for all particles (i.e., all agents).  The parameters are directly linked to a few general rules of interaction and decision and, importantly, our ambition is not to use them for a fit but rather we would like to be able to assign the value of the parameters a priori, based on socio-economical knowledge.  Admittedly, our model's number of parameters is still not small compared to models typically employed in time-series analysis, but keeping in mind our ambition, it just represents the complex nature of the problem.\\
\indent On the contrary, in the agent-based simulation approach, the number of parameters is huge, roughly proportional to the number of agents\footnote{\textcolor{black}{Notice that the number of \textit{Sinus-Milieus}$^{\tiny{\text{\textregistered}}}$ categories in Appendix \ref{app:SinusMilieus} that for folklore can be described by our model are only five and indistinguishable with respect to the rules of interaction among them and with the environment. In \cite{PSM2015TFSS}, where an agent-based simulation is adopted, the authors consider eight \textit{Sinus-Milieus}$^{\tiny{\text{\textregistered}}}$ categories and a corresponding set of parameters.}\label{footnoteSM}}. This is reminiscent of what is done in Physics, where one could fit the available data with an empirical law (the opposite of what we do in the present paper), but alternatively, one could also develop a theory that understands the available data and allows for generalization and prediction (our main aim in the present work). In particular, our purpose is to understand the link between the ``microscopic" dynamics of interacting agents and the time series of PV adoption, which represents, from our natural point of view, the cumulative result of a collective behaviour.\\
\indent Another difference between our approach and the agent-based simulation one is that it accommodates a sequential description of the process that leads an agent to adopt a PV, thus explaining the human psychology on the theme of GET. Notice that this sequential description catches the actual behaviour declared by adopters in response to surveys; see, e.g., \cite{KVM2017}.\\
\indent Finally, the present paper extends in a by far non-trivial way the model proposed in  \cite{FCLLMP2022ArXiv} by explicitly characterizing the transition rate from one state to another. This allows us to discuss some possible policy scenarios, such as a scenario in which we modify investment costs, a scenario in which the government supports photovoltaics, a scenario in which a nudging strategy is implemented, and a scenario in which social interaction is strengthened. In particular, our numerical simulations contribute to put in evidence the role of some key parameters. They can indicate how to devise external actions to eventually obtain a specific behaviour of the society under consideration.}

\subsection{Organization of the paper}
In Section \ref{sec:MarkovianModels}, we describe the two Markovian models for the GET. Section \ref{sec:numerical_experiments} describes the model's calibration, whereas the policy scenarios are discussed in Section \ref{sec:scenarioanalysys}. \textcolor{black}{Section \ref{sec::ethical_and_policy_implications} discusses the ethical and policy implications of the present article. Finally, Section \ref{sec:conclusions} presents the article's conclusions. Appendix \ref{app:investment_costs_and_cash_flows} describes how to compute the NPV, whereas Appendix \ref{app:SinusMilieus} present the so-called Sinus-Milieus$^{\tiny{\text{\textregistered}}}$ characterization.}

\section{Markovian models for the Green Energy Transition}\label{sec:MarkovianModels}
\begin{figure}
    \centering
    \includegraphics[scale=0.45]{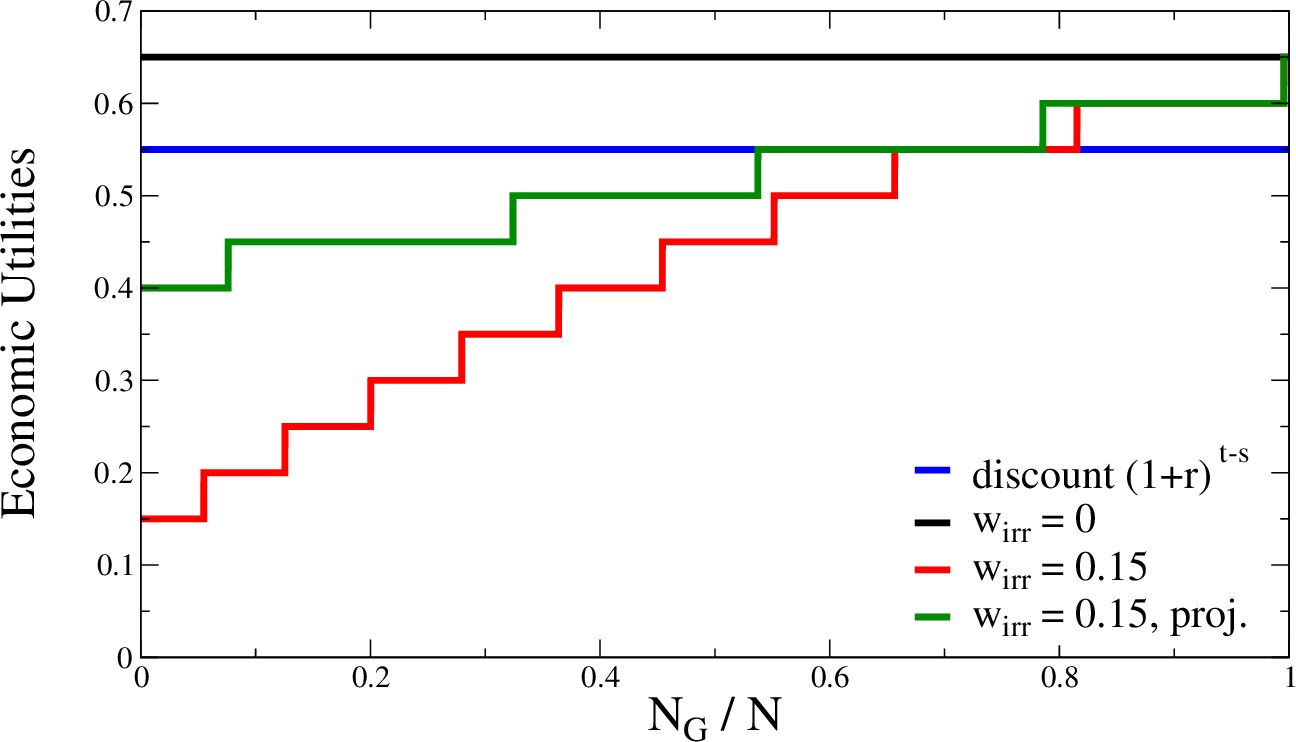}
    \caption{Pictorial representation of the economic utility in Equation \eqref{eq:econ} as a function of $n_G(t):=\frac{N_G(X_t)}{N}$ for a fixed $t$ for different discount factors $g^{i}(t, s, n_G(t))$ as defined in Equation \eqref{eq:discountfactor}.}
    \label{fig:npvs}
\end{figure}

This section details the two Markovian models we have briefly described in the introduction. In Section \ref{sec:numerical_experiments}, we will use only the second model, but \textcolor{black}{since the latter is a refinement of the first one,} we find it pedagogical to present both the models here.\\
\indent In the first model, we consider a world in which $N$ agents are characterized by a state variable at time $t$, say $X_t^{i}$, $i \in \{1, \ldots, N\}$, which can take one of the following three qualitative values: $X_t^{i} \in \{C, I, G\}$; and by a vector of random weights, denoted by $(w^i_{\text{ec}}, w^i_{\text{soc}}, w^i_{irr})$, that characterizes the individual in
several aspects; $w^i_{\text{ec}}, w^i_{\text{soc}}, w^i_{irr}$ are random variables distributed according to a triangular distribution. For simplicity, we also assume that these weights are independent. The state $C$ means that agent $i$ is ``Carbon". This expression is (admittedly) very vague. It indicates agents that can be ignorant, with a lack of awareness and limited thinking about the problem of GET. However, otherwise, they are prone to change their mind by gathering information from different external resources. We here count on three different resources: (I) We count on the neighbours, relatives, friends, and co-workers to pass information via ``word of mouth" to help spread energy efficiency, interest, and advantages. (II) We count on advertising and public education campaigns. (III) We count on a social utility, representing the comfort given by the impact of the agent's action on society. The model assumes that once the agent has been acquainted with (I), (II), and (III), she/he will make the transition from the state ``Carbon" to the state $I$, which stands for ``Informed." We assume that the rate of transition, denoted by $\lambda_{N}^{\textit{C}\rightarrow\textit{I}}$, depends on a quantity related to (I), a quantity related to (II), and a quantity related to (III). The former is given by a function of the fraction of ``Greens" at time $t$. The second one, if we consider a feedback development in communication, is also a function of the fraction of ``Greens" at time $t$. According to the following reasoning, the latter is also a function of the fraction of ``Greens" at time $t$. If $\frac{N_G(t)}{N}$ is small, then the impact of the individual $i$ is almost irrelevant since she/he feels that her/his choice is not a social phenomenon. On the other hand, the impact of the individual increases with $\frac{N_G(t)}{N}$ since she/he feels that her/his choice is beneficial for society. In conclusion, the dependence on these three factors can be summarised as the dependence on the ratio $\frac{N_G(t)}{N}$ and the number of green agents in agent $i$'s network. Assuming that the influence occurring locally is somewhat representative of that occurring globally, we conclude that the dependence on these three factors can be summarised as the dependence on the ratio $\frac{N_G(t)}{N}$. This makes our model a mean-field model. Formally, let $\mathcal{S} = \{C, I, G\}$ and $X_t = (X^{1}_t,\ldots, X^{N}_t) \in \mathcal{S}^{N}$ a generic configuration. We denote by $\frac{N_{G}(X_t)}{N}$ the fraction of ``Greens" at time $t$, where $N_{G}(X_t):=\sum_{j=1}^{N}\mathds{1}_{\{X_t^{j}=G\}}$ is the number of ``Greens" at time $t$. The probability to pass from $C$ to $I$ in a time interval $\Delta t \rightarrow 0$ is therefore defined in the following way:
\begin{equation}\label{eq:probItoC}
\begin{split}
    &\text{Prob}(X_{t+\Delta t}^{i} = I | X_{t}^{i} = C) := \lambda_{N}^{C \rightarrow I}\left(w_{\text{soc}}^{i}, \frac{N_{G}(X_t)}{N}\right)\cdot\Delta t\\
    &\quad\quad\quad\text{where}\,\,\lambda_{N}^{C \rightarrow I}\left(w_{\text{soc}}^{i}, \frac{N_{G}(X_t)}{N}\right) := w_{\text{soc}}^{i} \cdot F\left(\frac{N_{G}(X_t)}{N}\right).
\end{split}
\end{equation}
\noindent In the previous equation, $w_{\text{soc}}^{i}$ is a positive random variable taking values in the unit interval and indicating how much the agent is influenced by the three resources described above, the symbol $``:="$ means ``defined as" and the function $F:\,[0,1]\rightarrow[0,1]$ is, e.g., the identity function; see Section \ref{sec:numerical_experiments}. At this point, the model assumes that a barrier that prevents the agent from implementing the energy efficiency project (i.e., installing the PVs) is costs/uncertainty about payback. In particular, we assume that the probability of passing from $I$ to ``Greens", denoted by $G$, in a time interval $\Delta t \rightarrow 0$ is defined in the following way: \begin{equation}\label{eq:probGtoI}
\begin{split}
    &\text{Prob}(X_{t+\Delta t}^{i} = G | X_{t}^{i} = I) := \lambda_{N}^{I \rightarrow G}\left(w_{\text{ec}}^{i}, w_{\text{irr}}^{i}, \frac{N_{G}(X_t)}{N}\right)\cdot\Delta t\\
    &\quad\quad\quad\text{where}\,\,\lambda_{N}^{I \rightarrow G}\left(w_{\text{ec}}^{i}, w_{\text{irr}}^{i}, \frac{N_{G}(X_t)}{N}\right) = w_{\text{ec}}^{i} \cdot U_{\text{ec}}\left(w_{\text{irr}}^{i}, \frac{N_{G}(X_t)}{N}\right).
\end{split}
\end{equation}
\noindent In the previous equation, $w_{\text{ec}}^{i}$ denotes the importance that agent \textit{i} gives to the economic utility $U_{\text{ec}}$; the latter depends upon $w_{\text{irr}}^{i}$ which captures the bounded rationality of the agent and on $\frac{N_G(X_t)}{N}$, i.e., the fraction of ``Greens" at time $t$. For the computation  of the economic utility, we take inspiration from \cite{PSM2015TFSS}. We define it in the following way (the explicit dependence on $w_{\text{irr}}^{i}$ and $\frac{N_{G}(X_t)}{N}$ will be detailed below):
\begin{equation}\label{eq:econ}
\hspace{-0.2cm}
    U_{\text{ec}}\left(w_{\text{irr}}^{i}, \frac{N_{G}(X_t)}{N}\right) = \frac{\max(pp)-pp(i)}{\max(pp)-\min(pp)} = \frac{21-pp(i)}{20},
\end{equation}
where $pp(i)$ is the so called \emph{payback period} (or \emph{payback time)} of a specific PV system for agent \textit{i}. The payback period is determined by the year in which the Net Present Value (\text{NPV}, henceforth) of the PV system turns from negative to positive. More in detail, the \text{NPV} at time $t$ is defined as:
\begin{equation}\label{eq:netpresentvalue}
    \text{NPV}(t,n_G(t), \tau) = - I_{\text{econ}}(t) + \sum_{s = t+1}^{\tau} \frac{R(s-t)}{1 + g^{i}(t, s, n_G(t)))}, \quad t\leq\tau\leq t+20,
\end{equation}
and it depends on the fraction of ``Greens" at time $t$, $n_G(t):=\frac{N_{G}(X_t)}{N}$\footnote{Henceforth, we will use interchangeably the two notations.} via the discount factor $g^{i}(t, s, n_G(t)))$. The discount factor is agent-specific, and it is defined as:
\begin{equation}\label{eq:discountfactor}
    g^{i}(t, s, n_G(t))) := w^{i}_{\text{irr}} \cdot (1-n_G(t)) \cdot (s-t).
\end{equation}
We now describe the quantities in Equation \eqref{eq:netpresentvalue} and discuss later the discount factor in Equation \eqref{eq:discountfactor}. $I_{\text{econ}}(t)$ are the investment costs. Instead, the cash flow $R(s)$ comprises five factors. The term $R_{\text{Save}}(s, \text{CE})$ includes all earnings that are generated by directly using the produced electricity instead of buying it from or selling it to the grid operator. The terms $R_{\text{Gov}}(s, \text{CE})$, $R_{\text{Adm}}(s)$, $R_{\text{Main}}(t)$, $R_{\text{Deprec}}(s)$ and $R_{\text{time}}(s)$ indicate cash flows due to governmental support, administrative fees, maintenance and upfront costs, depreciation allowance payments, and the cash equivalent of the time spent for the administrative consultancy. 
\begin{equation}\label{eq:cash_flows}
\hspace{-0.15cm}
    R(s) = R_{\text{Save}}(s, \text{CE}) + R_{\text{Gov}}(s, \text{CE}) - R_{\text{Adm}}(s, \text{CE}) - R_{\text{Main}}(s) - R_{\text{Deprec}}(s)-R_{\text{time}}(s),
\end{equation}
where \text{CE} stands for Conto Energia (\textcolor{black}{see Appendix \ref{app:investment_costs_and_cash_flows}}). Since the cash flow computation is not our contribution, we confine its description in Appendix \ref{app:investment_costs_and_cash_flows}. Notice that the state $G$ is absorbing, in the sense that an agent may jump from the state $I$ to the state $G$ but cannot jump back from $G$ to $I$, or $C$.\\
\indent In the model we have just presented, once the agent is informed, she/he evaluates the economic utility and passes from $I$ to $G$ with a rate that is proportional to the latter. 
In the second model, we propose a more detailed description of the procrastination loop in which an agent may end up trapped due to the inter-temporal structure of the green choice.   \textcolor{black}{Before proceeding, we observe that an individual qualifies as a procrastinator for an action, say $A$, at time $t$ if the following conditions are given:
\begin{itemize}
    \item[(i)]  at time $t$, the individual prefers to postpone $A$ to a time $t+T$;
    \item[(ii)] at time $t-T$, the individual prefers to perform $A$ no later than time $t$;
    \item[(iii)] at time $t+T$, the individual regrets not having performed $A$ earlier.
\end{itemize}}
In order to gain \textcolor{black}{the previous mechanism}, we propose to extend the number of qualitative values that the variable $X_t^{i}$ can assume.
In this second model, indeed, $X_t^{i} \in \{C, I, PL, G\}$. The states $\{C, I, G\}$ have the same meaning as before. The state $PL$, instead, stands for ``Planner"; it indicates an agent that has acquired sufficient information on the benefits of PVs, or has developed a certain level of sensitivity on climate change and environmental issues, and \textit{plans} to install the PVs \textcolor{black}{by looking at the ``projected in the future" economic utility. This last concept is new and we will clarify it in the next few lines. The introduction of the new state “Planner” aims to capture the following behaviour. In general, it is not enough that an individual has developed a clear preference for purchasing a PV system to pass from ``Informed" to ``Green" because the value of initial investment costs, previously judged to be lower than that of future energy benefits, suddenly becomes higher.} Precisely, the agent evaluates the latter and passes from $I$ to $PL$ in a time interval $\Delta t \rightarrow 0$ according to the following probability
\begin{equation}\label{eq:probItoPL}
\begin{split}
    &\text{Prob}(X_{t+\Delta t}^{i} = PL | X_{t}^{i} = I) := \lambda_{N}^{I \rightarrow PL}\left(w_{\text{ec}}^{i}, w_{\text{irr}}^{i}, T, \frac{N_{G}(X_t)}{N}\right)\cdot\Delta t\\
    &\quad\quad\quad\text{where}\,\,\lambda_{N}^{I \rightarrow PL}\left(w_{\text{ec}}^{i}, w_{\text{irr}}^{i}, T, \frac{N_{G}(X_t)}{N}\right) = w_{\text{ec}}^{i} \cdot U_{\text{ec}}^{proj}\left(w_{\text{irr}}^{i}, T, \frac{N_{G}(X_t)}{N}\right).
\end{split}
\end{equation}
In the previous equation, $U_{\text{ec}}^{proj}$ is defined as in Equation \eqref{eq:econ} in which the \text{NPV} at time $t$ is given by:   
\begin{equation}\label{eq:NPVItoPL}
    \text{NPV}(t, t+T, n_{G}(t), \tau) = - \frac{I_{\text{econ}}(t+T)}{1+g^{i}(t, t+T, n_G(t))} + \sum_{s=t+T}^{\tau}\frac{R(s-(t+T-1))}{1+g^{i}(t,s,n_G(t))}
\end{equation}
Finally, the probability of passing from $PL$ to $G$ coincides with the probability in Equation \eqref{eq:probGtoI}, with $I \equiv PL$. \textcolor{black}{The Markovian dynamics described above capture the procrastinator traits outlined in (i)-(iii). Let's say an individual is in the ``Informed'' state at time \( t-T \). Based on the probability defined in \eqref{eq:probItoPL}, the individual can either transition to the ``Planner'' state or remain in the ``Informed'' state. If the first event occurs, then she/he plans to install PV no later than time $t$ (condition (ii)). At time $t$, we may observe either that the individual has switched to the ``Green" state or that she/he has remained a ``Planner''. If the latter case happens, it indicates that the individual has postponed the action of installing PV. This is because, even though the expected future economic benefit is favorable, the current evaluation of the economic benefit is not (condition (i)).}
At this point, the following observations are in order. Figure \ref{fig:npvs} displays the economic utility in Equation \eqref{eq:econ} as a function of $n_G(t)=\frac{N_G(X_t)}{N}$ for a fixed $t$ when using four different discount factors. Each discount factor can be thought to correspond to four different types of agents. The economic utility in black corresponds to a \text{NPV} in Equation \eqref{eq:netpresentvalue} where $g^{i}(t, s, n_G(t))$ is equal to zero. It does not depend, as expected, on $n_G(t)$: this matches a discount factor of an agent that does behave neither like a \textit{homo economicus} nor like an agent that is bounded rational. Indeed, agents that think and behave like \textit{homo economicus} would discount each addend in Equation \eqref{eq:netpresentvalue} by $(1+r)^{t-s}$, where $r$ is the interest rate; the economic utility for such agents is displayed in blue. Again, the latter does not depend, as expected, on $n_G(t)$. Instead, agents that are bounded rational would discount each addend in Equation \eqref{eq:netpresentvalue} by $1+g^{i}(t, s, n_G(t))$, where $g^{i}(t, s, n_G(t))$ is defined as in Equation \eqref{eq:discountfactor}. The corresponding utility function is displayed in red. First, we observe that this utility is increasing with respect to the ratio of green agents. This represents the fact that when the number of green agents increases, the social pressure is higher and the effect of the hyperbolic discount factor is softened.
Finally, notice that $U_{\text{ec}}^{proj}\left(w_{\text{irr}}^{i}, T, \frac{N_{G}(X_t)}{N}\right)$ is higher that the corresponding $U_{\text{ec}}\left(w_{\text{irr}}^{i}, \frac{N_{G}(X_t)}{N}\right)$ \textcolor{black}{and that the reversal of preferences is temporary since an ``Informed" can become ``Green”, as in all cases of procrastination (see the qualification as a procrastinator defined in (i)--(ii) above).} In conclusion, we highlight that increasing the value of $w^{i}_{\text{irr}}$ would decrease the economic utilities. Indeed, the higher $w^{i}_{\text{irr}}$, the bigger the misperception of such a utility function.

\textcolor{black}{
\subsection{Mean Field derivation of the individual based model}\label{subsec:macroscopic_derivation}
We conclude this section by deriving the so-called mean field limit of our second model. As explained in Subsection \ref{subsec:LiteratureAndContribution}, our system of agents can be considered a system of interacting particles, where the particles are all equal and subject -- more or less -- to the same rules of interaction among them and with the environment. The mean field is the general method that allows to summarize the behaviour of the particles in a few ``macroscopic" laws and equations; most of the numerical simulations are based on these macroscopic laws.\\
\indent Let $X_t = (X_t^1, \ldots, X_t^N) \in \mathcal{S}$, with $\mathcal{S}$ now defined as $\mathcal{S}:=\{C, I, PL, G\}$, be a generic configuration, $w_{\text{soc}}$, $w_{\text{irr}}$ and $w_{\text{ec}}$ be the weights defined above, and $\mathcal{P}_2(\mathcal{S}\times\mathbb{R}^3)$, $\mathcal{P}_2(\mathcal{S})$, and $\mathcal{P}_2(\mathbb{R})$ be the space of probability measures (on the corresponding spaces) that are square integrable. We suppose that the weights are constant over time once the simulation starts, but they are also sampled from a distribution at time zero. In what follows, we will denote by capital letters the corresponding random variables. At this point, we can define the following quantities:
\begin{equation}
\begin{split}\label{eq::nquantities}
    &\nu_t^{N}:= \frac1N \sum_{i=1}^N\delta_{\left(X^i_t, w^i_{\text{soc}},w^i_{\text{ec}},w^i_{\text{irr}}\right)}\in \mathcal{P}_2\left(\mathcal S\times \mathbb{R}^3\right)\quad\,\, \mu^{N}_t:=\frac1N \sum_{i=1}^N\delta_{X^i_t}\in \mathcal{P}_2\left(\mathcal S\right)\\
    &f^N_{\text{soc}}(dw_{\text{soc}}):=\frac1N \sum_{i=1}^N\delta_{w^i_{\text{soc}}}\in\mathcal{P}_2\left(\mathbb{R}\right)\quad\quad\quad f^N_{\text{ec}}(dw_{\text{ec}}):=\frac1N \sum_{i=1}^N\delta_{w^i_{\text{ec}}}\in\mathcal{P}_2\left(\mathbb{R}\right)\\
    &\quad\quad\quad\quad\quad\quad\quad\quad f^N_{\text{irr}}(dw_{\text{irr}}):=\frac1N \sum_{i=1}^N\delta_{w^i_{\text{irr}}}\in\mathcal{P}_2\left(\mathbb{R}\right)\\\
    &n^{C,N}(X_t):=\frac1N \sum_{i=1}^N \mathds1_{\{X^i_t=C\}}\in (0,1)\quad\quad\quad\,\, n^{I,N}(X_t):=\frac1N \sum_{i=1}^N \mathds1_{\{X^i_t=I\}}\in (0,1)\\ 
    &n^{PL,N}(X_t):=\frac1N \sum_{i=1}^N \mathds1_{\{X^i_t=PL\}}\in (0,1)\quad\quad\,\,\, n^{G,N}(X_t)=\frac1N \sum_{i=1}^N \mathds1_{\{X^i_t=G\}}\in (0,1).\\ 
\end{split}
\end{equation}
\noindent   The goal is to find an expression for the evolution of $n^{C,N}(X_t)$, $n^{I,N}(X_t)$, $n^{PL,N}(X_t)$ and $n^{G,N}(X_t)$. Toward this aim, we make the following assumption. We assume that the weights are independent from each other and independent from $X_t^{i}$ for each fixed $t$, i.e.:
\begin{equation*}
    \nu_t^{N}(dx, dw_{\text{soc}}, dw_{\text{ec}}, dw_{\text{irr}}) = \mu^{N}(dx)\cdot f^N_{\text{soc}}(dw_{\text{soc}}) \cdot f^N_{\text{ec}}(dw_{\text{ec}})\cdot f^N_{\text{irr}}(dw_{\text{irr}}).
\end{equation*}
Now, we consider an observable $F\,:\,\mathcal{S} \rightarrow \mathbb{R}$ and $X:=(X^{1},\ldots, X^{N}) \in \mathcal{S}$. The process $X_t = (X_t^{1},\ldots,X_t^{N})$ is a continuous-time Markov chain (of cellular automaton type) with the following time-dependent infinitesimal generator:
\begin{equation*}\label{eq:generator}
\begin{split}
\mathcal{L}_t F(X) &= \sum_{i=1}^N\mathds1_{\{X^i=C\}} \lambda_{N}^{C\to I}\left(w^i_{\text{soc}}, n^{G,N}(X)\right)\left(F(X^{i,C\to I})-F(X)\right)\\
                   &+ \sum_{i=1}^N\mathds1_{\{X^i=I\}} \lambda_{N}^{I\to PL}\left(w^i_{\text{ec}}, w^i_{\text{irr}}, n^{G,N}(X)\right)\left(F(X^{i,I\to PL})-F(X)\right)\\
                   &+ \sum_{i=1}^N\mathds1_{\{X^i=PL\}} \lambda_{N}^{PL\to G}\left(w^i_{\text{ec}}, w^i_{\text{irr}}, n^{G,N}(X)\right)\left(F(X^{i,PL\to G})-F(X)\right).\\
\end{split}
\end{equation*}
At this point, we can use the previous expression to time-dependent generators $\mathcal{L}_t n^{C,N}(X)$, $\mathcal{L}_t n^{I,N}(X)$, $\mathcal{L}_t n^{PL,N}(X)$ and $\mathcal{L}_t n^{G,N}(X)$, where $n^{C,N}(X)$, $n^{I,N}(X)$, $n^{PL,N}(X)$ and $n^{G,N}(X)$ are defined in Equation \eqref{eq::nquantities}; one has to replace the observable $F$ with $n^{C,N}(X)$, $n^{I,N}(X)$, $n^{PL,N}(X)$ and $n^{G,N}(X)$, respectively. We compute explicitly $\mathcal{L}_t n^{C,N}(X)$ since the others can be derived by using a similar argument.
\begin{equation*}
    \begin{split}
        \mathcal{L}_t n^{C,N}(X) &= \sum_{i=1}^N\mathds1_{\{X^i=C\}} \lambda_{N}^{C\to I}\left(w^i_{\text{soc}}, n^{G,N}(X)\right)\left(n^{C,N}(X^{i,C\to I})-n^{C,N}(X)\right)\\
                                 &+ \sum_{i=1}^N\mathds1_{\{X^i=I\}} \lambda_{N}^{I\to PL}\left(w^i_{\text{ec}}, w^i_{\text{irr}}, n^{G,N}(X)\right)\left(n^{C,N}(X^{i,I\to PL})-n^{C,N}(X)\right)\\
                                 &+\sum_{i=1}^N\mathds1_{\{X^i=PL\}} \lambda_{N}^{PL\to G}\left(w^i_{\text{ec}}, w^i_{\text{irr}}, n^{G,N}(X)\right)\left(n^{C,N}(X^{i,PL \to G})-n^{C,N}(X)\right)\\
                                 &=-\frac{1}{N}\sum_{i=1}^{N} \mathds1_{\{X^i=C\}}\lambda_{N}^{C\to I}\left(w^i_{\text{soc}}, n^{G,N}(X)\right)\\
                                 &=- \int_{\mathbb{R}} w_{\text{soc}}\cdot n^{G, N}(X)\cdot f_{\text{soc}}^{N}(d w_{\text{soc}}) \cdot n^{N,C}(X)\\
                            &=- \mathbb{E}^{N}\left[W_{\text{soc}}\right]\cdot n^{N,C}(X)\cdot n^{N,G}(X),
    \end{split}
\end{equation*}
where in the penultimate equality we use the rewriting of the term in terms of the measure $\nu^{N}$ and the independence between the random variables. In addition, the time-dependent infinitesimal generators $\mathcal{L}_t n^{I, N}$, $\mathcal{L}_t n^{PL, N}$ and $\mathcal{L}_t n^{G, N}$ are given by:
\begin{eqnarray*}
\hspace{0cm}
    \begin{split}
        \mathcal{L}_t n^{I,N}(X) &= \mathbb{E}^{N}\left[W_{\text{soc}}\right]\cdot n^{C,N}(X)\cdot n^{G,N}(X)\\ 
        &-\mathbb{E}^{N}\left[W_{\text{ec}}\right]\cdot\mathbb{E}^{N}[U(W_{\text{irr}},n^{N,G}(X))]\cdot n^{I,N}(X).\\
        \mathcal{L}_t n^{PL,N}(X) &= \mathbb{E}^{N}\left[W_{\text{ec}}\right]\cdot \mathbb{E}^{N}[U(W_{\text{irr}},n^{G,N}(X))]\cdot n^{I,N}(X)\\
                                  &-\mathbb{E}^{N}\left[W_{\text{ec}}\right]\cdot\mathbb{E}^{N}[U(W_{\text{irr}},n^{N,G}(X))]\cdot n^{PL,N}(X).\\
        \mathcal{L}_t n^{G,N}(X) &= \mathbb{E}^{N}\left[W_{\text{ec}}\right]\cdot \mathbb{E}^{N}[U(W_{\text{irr}},n^{G,N}(X))]\cdot n^{I,N}(X).
    \end{split}
\end{eqnarray*}
Finally, by using the It\^o-Dynkin Equation, the second It\^o-Dynkin Equation (see \cite{KL1998}, Appendix A), and by taking the limit for $N \rightarrow \infty$, we obtain the following final system of ordinary differential equations: 
\begin{equation}\label{eq:mf_approximation}
\begin{cases}
 \frac{d}{dt}n^{C}(X)=-\mathbb{E}\left[W_{\text{soc}}\right]\cdot n^{C}_t(X)\cdot n^{G}_t(X)\\
  \frac{d}{dt}n^{I}_t(X)=\mathbb{E}\left[W_{\text{soc}}\right]\cdot n^{C}_t(X)\cdot n^{G}_t(X)\\
\quad\quad\quad\quad\quad -\mathbb{E}\left[W_{\text{ec}}\right]\cdot \mathbb{E}\left[U^{\text{proj}}_{\text{ec}}\left(W_{\text{irr}}, T, n^{G}_t(X)\right)\right]n^{I}_t(X)\\
\frac{d}{dt}n^{PL}_t(X)=\mathbb{E}\left[W_{\text{ec}}\right]\cdot \mathbb{E}\left[U^{\text{proj}}_{\text{ec}}\left(W_{\text{irr}}, T,n^{G}_t\right)\right]\cdot n^{I}(X)\\
\quad\quad\quad\quad\quad -\mathbb{E}\left[W_{\text{ec}}\right]\mathbb{E}\left[U_{\text{ec}}\left(W_{\text{irr}}, n^{G}_t(X)\right)\right] n^{PL}_t(X)\\
\frac{d}{dt}n^{G}_t(X)=\mathbb{E}\left[W_{\text{ec}}\right]\cdot \mathbb{E}\left[U_{\text{ec}}\left(W_{\text{irr}}, n^{G}_t(X)\right)\right]\cdot n^{PL}_t(X).\\
 \end{cases}
\end{equation}
As said, we will use it in most of our numerical simulation. Before proceeding, the following important remark on the weights is in order.  
\begin{remark}\label{rmk:remarkthree}
The dependence on the weights $w_{\text{ec}}^{i}$ and $w_{\text{soc}}^{i}$ is linear in the transition rates, whereas the dependence on $w_{\text{irr}}^{i}$ is non-linear. In addition, the distribution matters and it appears in the term $\mathbb{E}\left[U_{\text{ec}}\left(W_{\text{irr}}, n^{G}_t(X)\right)\right]$.   
\end{remark}}

{\color{black}\section{Numerical experiments}\label{sec:numerical_experiments}
In this section, we first present the data we will use in our analysis; see Subsection \ref{subsec:datadescription}, and the model's inputs, such as the cash flows in Equation \eqref{eq:cash_flows}; see Subsection \ref{subsec:mode_inputs}. Then, we describe the choice of the number of agents $N$ employed in the simulations; see Subsection \ref{subsec:number_agents}. Finally, Subsection \ref{subsec:modelcalibration} presents the model's calibration, along with a thorough stability and sensitivity analysis.
\subsection{Data description}\label{subsec:datadescription}
In our analysis, we examine the number of installed PV systems in Italy from 2006 to 2020 using data from the GSE report \cite{GSE}. We focus on the proportion of homeowners who have adopted PV systems during this period. We make the following considerations: From 2010 to 2020, we have specific values for the ``Domestic" time series. For a detailed description of the four categories into which the national data is divided, please refer to Subsection \ref{subsec:MotivationAndModelling}. In 2009, due to a different categorization, we estimate that ``Domestic" corresponds to a certain percentage of the total installed PV systems. For the time period 2006-2008, we lack a specific percentage. However, given that residential PVs account for a substantial portion, we consider the total number of installed PV systems. These systems typically range between 1 and 20 kW, which aligns with choices made by homeowners. Additionally, we conduct an analysis of regional data. We specifically focus on regions where the percentage of ``Domestic" installations is notably high: Liguria, Friuli Venezia Giulia, and Veneto. In Table \ref{tab:averagepercentage}, we present the average percentage of ``Domestic" installations relative to the total power (\text{Avg.\,$\%$}) from 2010-2020 for these regions.
\begin{table}[h!]
\begin{center}
\begin{tabular}{|c|c|}
\hline
Region & \text{Avg.\,$\%$\,2010-2020\,PVs\,``Domestic"}\\ \hline
Liguria  & $23,48$                                                             \\ \hline
F. V. Giulia   & $23,29$                                                             \\ \hline
Veneto   & $19,05$                                                             \\ \hline
\end{tabular}
\end{center}
\caption{Avg.\,$\%$ of ``Domestic" PVs installation with respect the power over the period 2010-2020. \textit{Data Source\,:\,} GSE (\url{https://www.gse.it/dati-e-scenari/statistiche}).}
\label{tab:averagepercentage}
\end{table}
\begin{table}[h!]
\begin{center}
\begin{tabular}{|c|c|}
\hline
Region or country & \text{Ratio}\\ \hline

Italy  &  $0,73$                                                           \\ \hline
Liguria  & $0,61$                                                             \\ \hline
F. V. Giulia   & $0,88$                                                             \\ \hline
Veneto   & $3,80$                                                             \\ \hline
\end{tabular}
\end{center}
\caption{Ratio between the number of inhabited buildings and the number or families. \textit{Data Source\,:\,} \url{https://dati-censimentopopolazione.istat.it/}.}
\label{tab:ratio}
\end{table}\\
\subsection{Model's inputs}\label{subsec:mode_inputs}
\indent Here, we describe our model inputs. First, we must compute the cash-flows in Equation \eqref{eq:cash_flows}. We start from the term $I_{\text{econ}}(t)$ in Equation \eqref{eq:Iecon}; see Appendix \ref{app:investment_costs_and_cash_flows}. The authors in \cite{MRZ} indicate the numerical values for the latter quantity. The index ``Plant Cost compared to Modules Cost"  (for the crystalline silicon) can be considered equal to a value between 1.5 and 1.9; the plant size is 3 \textrm{kW}. The previous computation gives a result comparable with the ``Turnkey PV system" (residential) average prices obtained from the National Survey Report of PV Power Applications in Italy. The evolution of the price per installed \textrm{W} of a PV system over the period 2007-2020 is displayed in Figure \ref{fig:prices}. Second, we need to recover the value for $E_{\text{PV}}(s) = E_{\text{Sun}} \cdot P_{\text{MPP}} \cdot (1-\xi_{\text{Abrasion}})^{s-t-1}$. Admittedly, we were not able to find in \cite{PSM2015TFSS} and references therein a value for the coefficient of abrasion $\xi_{\text{Abrasion}}$. Therefore, we propose to use the following procedure. We define the value $E_{\text{PV}}(t)$ as the average amount of electricity generated by a household PV system located in Milano, Pisa, and Palermo, respectively. (\textit{Data Source\,:\,} \url{https://re.jrc.ec.europa.eu/pvg_tools/en/#api_5.1}):  
\begin{equation*}
    E_{\text{PV}}(t) = 3\cdot\frac{(1310.32+1397.34+1523.02)}{3}\quad\textrm{kWh},
\end{equation*}
where the 3 in front of the equation indicates that we are considering PV system with a size of 3 \textrm{kW}. Then we assume that $E_{\text{PV}}(t)$ decreases by $3\%$ every year. With this datum we can then compute $R_{\text{Save}}(s, \text{CE5})$ by choosing $\chi_{DC}=0.85$, $p_{\text{elec,buy}}=0.18475$ \textrm{Euro/kWh}, $p_{\text{elec,sell}}=0.06056$ \textrm{Euro/kWh}, $\tau_{\text{elec,buy}}=0.04302$, $\tau_{\text{elec,sell}}=0.03211$. As we are not interested in an exact computation, we consider the electricity prices constant in the simulation. However, we checked that our model can still fit the data, with slightly different parameter values, if we consider electricity yearly medium prices. As regards as, instead, $R_{\text{Gov}}(s, \text{CE})$ it depends on the year at which the simulation starts because of the difference in the values of the \textrm{FiT} (see Appendix \ref{app:investment_costs_and_cash_flows}). At this point, we need to specify the negative cash flows. As regards $R_{\text{Adm}}(s, \text{CE})$, we follow \cite{PSM2015TFSS} and we set it equal to $3 \frac{\text{Euro}}{\text{kW} \cdot \text{year}}$ for all the \text{CE}. As regards $R_{\text{Main}}$, we set its by choosing $\alpha_{\text{upfront}}=0.010$ and $\alpha_{\text{Main}}=0.013$. Finally, $R_{\text{time}}(s)$ is set to the standard value of $200$ \textrm{Euro}.
\begin{figure}
    \centering
    \includegraphics[scale=0.45]{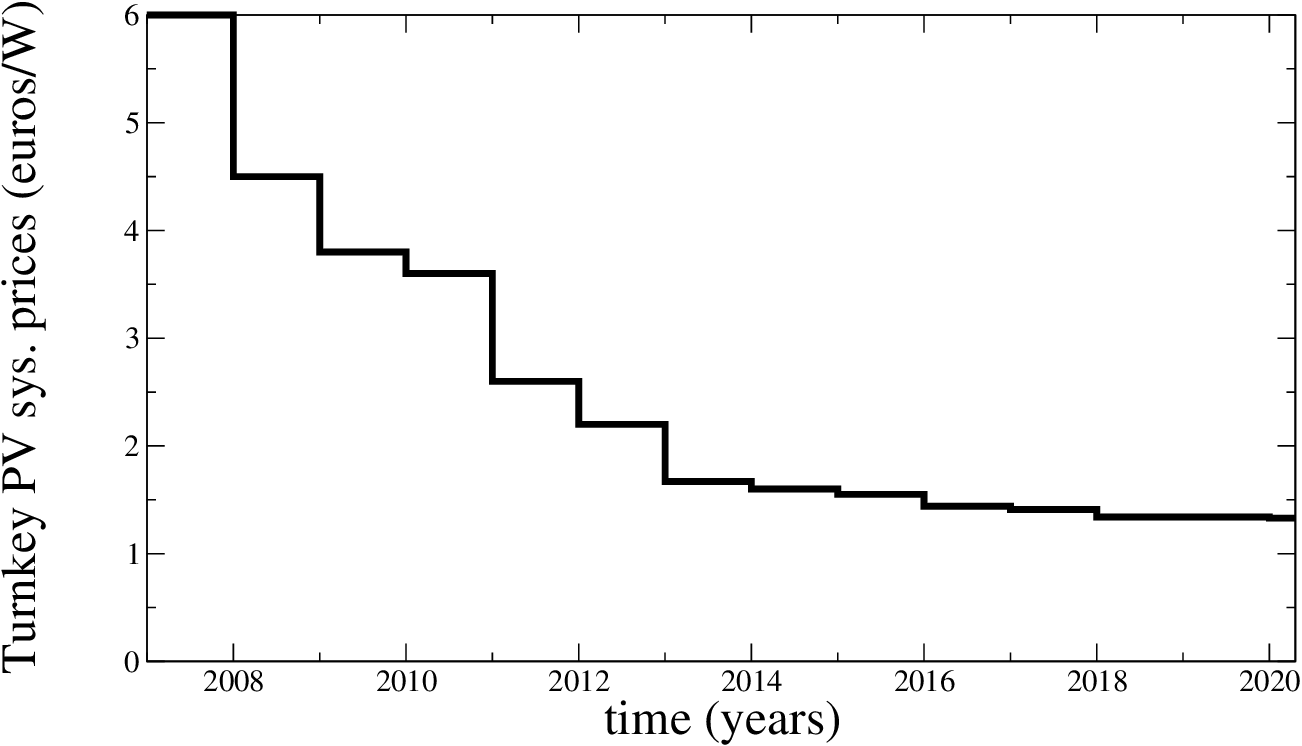}
    \caption{Turnkey PV system price per installed \textrm{W} of a PV system over 2007-2020, residential buildings. \textit{Data Source\,:\,} National Survey Report of PV Power Applications in Italy.}
    \label{fig:prices}
\end{figure}

\subsection{Fixing the number $N$ of agents in the system}
\label{subsec:number_agents}
Agents in our system are representative individuals of a small community, i.e., a family or an apartment building.  As a consequence,   we initially set $N$ to
\begin{equation}\label{eq:N}
        N \propto \max\{\# \text{of inhabited buildings}, \# \text{of families}\}.
\end{equation}
In order to fix $N$, we then compute the ratio between the number of inhabited buildings and the number of families (\textit{Data Source\,:\,} \url{https://dati-censimentopopolazione.istat.it/}).  Table \ref{tab:ratio} reports the ratio for Italy and the three considered regions. According to Equation \eqref{eq:N}, if ``Ratio" is less than one, then $N$ is set equal to the number of inhabited buildings. Instead, if the ``Ratio" is greater than one, then we fix $N$ as the number of families.\\
\indent Importantly, as pointed out in the footnote \footref{footnoteSM}, the number of \textit{Sinus-Milieus}$^{\tiny{\text{\textregistered}}}$ categories that for folklore can be described by our model are only five and indistinguishable with respect to the rules of interaction among them and with the environment. In particular, these socio-cultural categories include the population segment prone to the installation of PVs. E.g., the so-called {\em Tradizionali Conservatori} \textit{Sinus-Milieus}$^{\tiny{\text{\textregistered}}}$ category seems less inclined to consider this type of innovation; see \cite{PSM2015TFSS}.   We multiply $N$ by $\alpha=0.55$ to obtain the final number of agents in the system. It is important to note that the qualitative nature of the results remains robust regardless of this choice.  
\subsection{Model's calibration}\label{subsec:modelcalibration} 
The model's calibration is based on the indirect calibration approach commonly used in agent-based simulation. This involves running several simulations and comparing the results with empirical data to find the best model parameters. To evaluate the model's accuracy, the time series is divided into three equal parts. The first two-thirds is used to calibrate the model, and the remaining one-third is used to test the model's accuracy.\\ 
\indent In principle, we should calibrate the model concerning three weights: $w_{\text{ec}}$, $w_{\text{soc}}$, and $w_{\text{irr}}$, the time horizon for the ``projected in the future" economic utility (see Section \ref{sec:MarkovianModels}) $T$, and the three initial conditions for the system of ordinary differential equations in Equation \eqref{eq:mf_approximation}; we have a total of eight parameters. However, to alleviate a possible over-fitting problem, we calibrate the model only with respect to the three weights. Instead, we deduce $n_G(0)$ from real data, fix a priori $T$, $n_{I}(0)$ and $n_{P}(0)$, and derive $n_C(0)$ from a structural constraint. Table \ref{tab:par_model_simulation} summarizes what we have just said. To validate our procedure, we discuss in Subsection \ref{subsec:simulationsresults_stability} the stability of the model fitting with respect to the fixed parameters.

\begin{table}[htbp] 
    \centering
\begin{tabular}{lll} \toprule
        \textbf{Parameter} & \textbf{Description}& \textbf{Comment} \\ \midrule
        $w_{\text{ec}}$     & Expectation of the random variable $W_{ec}$, $\mathbb{E}\left[W_{ec}\right]$ & Calibrated\\
        $w_{\text{soc}}$    & Expectation of the random variable $W_{soc}$, $\mathbb{E}\left[W_{soc}\right]$ & Calibrated\\
        $w_{\text{irr}}$    & Expectation of the random variable $W_{irr}$, $\mathbb{E}\left[W_{irr}\right]$ & Calibrated\\
        $T$       & Time horizon ``projected in the future" utility & Fixed\\
        $n_I(0)$  &Initial condition of the ``Informed" population & Fixed\\
      $n_P(0)$  & Initial condition of the ``Procastinators" population& Fixed\\
    $n_G(0)$ & Initial condition of the ``Green" population& Real Data\\
    $n_C(0)$    &Initial condition of the ``Carbon" population & Derived from a \\
           & & structural constraint.\\
        \bottomrule
    \end{tabular}
    \caption{The table reports the parameters of the model that should be calibrated with the relative description in column \textit{Description}. Column \textit{Comment} indicates whether the parameter is calibrated, fixed a priory, deduced from real data or from a structural constraint.}
 \label{tab:par_model_simulation}
\end{table}

\noindent We now describe the choice of $n_P(0)$, $n_I(0)$, $n_G(0)$, $n_C(0)$, and $T$.  The initial number of ``Greens" agents, $n_G(0)$, coincides with the number of PV adopters in 2007. As regards the initial number of ``Informed" agents $n_I(0)$ and ``Planner" agents $n_P(0)$, it is reasonable to set them as being proportional to the initial number of ``Greens" and ``Procrastinator", respectively, i.e.,
\begin{equation*}
    n_P(0)=k_P n_G(0),\,\,\text{and}\,\,n_I(0)=k_I n_P(0),
\end{equation*}
with $k_P$ and $k_I$ greater than one; we set $k_P=k_I=10$. Instead, the initial number of ``Informed" is set by using the following structural constraint 
\begin{equation*}
    n_C(0)+n_I(0)+n_P(0)+n_G(0)=1.
\end{equation*}
\noindent   Finally, we set $T=5$. As said, the weights $w_{\text{ec}}$, $w_{\text{soc}}$ and $w_{\text{irr}}$ are calibrated via the indirect calibration approach by using as loss function the mean square error. Precisely, the resulting values for the weights $w_{\text{ec}}$, $w_{\text{soc}}$  determine the average of a triangular distribution with support over $[0,1]$; notice that one of the most common distributions adopted in this type of literature is the triangular one. In principle, we can also consider time-varying weights and draw a different realization of the weights at each time step. However, since we will consider a ``sufficiently" large number of agents in our numerical experiments, random weights $w^i_{\text{soc}}, w^i_{\text{ec}}$ will be replaced by their average; see the system in Equation \eqref{eq:mf_approximation}. The same simplification cannot be applied for the weight $w^i_{\text{irr}}$ because it appears hidden in a non-linear function.\\
\indent Alternatively, an interesting methodology is the one proposed in \cite{ZMCS2022SMPT} for the diffusion of PV systems in the Netherlands. They identify four \emph{factors} ((a) Advertising; (b) Neighborhood; (c) household income; (d) payback period of a PV system) and \emph{related to} some \emph{aspects} ((1) The contribution to a better natural environment; (2) The grant on offer; (3) The central organization of the request for a grant; (4) Independence from electricity supplier; (5) Discussion with other owners convinced me to adopt; (6) The buying of PV systems by neighbours/acquaintances; (7) The technical support offered by the municipality. To the latter, they assigned a score between 1 and 5 as in \cite{WJ2006EP}. Then, the resulting triangular distribution's support is $[1, 5]$, and the mean is obtained from the average of the score of the pair factors-aspects. Although very interesting, we will leave this type of approach for further research.\\
\indent Table \ref{tab:weights} reports the optimal value of the weights $w_{\text{soc}}, w_{\text{ec}}$, and $w_{\text{irr}}$; we also report the calibrated weights in Figure \ref{fig:histogram} for a nicer representation. At this point, we simulate both the agent-based system and its mean field limit equation to show that the latter is actually an approximation of the former.   More precisely, in the agent-based system case, we update the configurations on a monthly basis, and we display the sample mean; the computation of $\mathbb{E}\left[U_{\text{ec}}\left (W_{\text{irr}}, n^{G}_t(X)\right)\right]$ is performed with 10000 samples. Instead, the simulation of the mean field equation is performed with a time-step of $0.01$. In both cases, the cash flow $R$ (see Equation \eqref{eq:cash_flows}) is taken to be constant for the entire year. Figure \ref{fig:regions} shows the results of the simulation of the total number of adopters over the number of buildings in the chosen regions (top three sub-figures) and in Italy (bottom figure). The period fit range, i.e., the first two-thirds of the data, is displayed in the shade of blue. The diagrams illustrate the actual PV market data and the simulation of our model, which displays a very good fit for the actual number of adopters.\\ 
\indent The following observation regarding Figure \ref{fig:regions} is in order. There are three distinct phases. At the initial formation phase, high costs (see Figure \ref{fig:prices}) and uncertainty result in slow and erratic growth. This formative phase ends with a ``take-off," which kicks the growth phase, in which growth accelerates due to positive feedback in economic profitability, technology learning and governmental support via the different phases of the CE. After achieving its maximum level, growth begins to slow mainly because of the elimination of the incentives. Notice that we do not interpret, as in \cite{CVTGJ2021Nature}, this phase as a saturation phase. In particular, the growth phase after the initial formation period is captured mainly by the variation in the economic utility: indeed, in that period, the $N_{G}$ curve mirrors the one of the payback period, for which we report a typical pattern over 2007--2020 in Figure \ref{fig:pp}. For comparison, we also report the payback period for an \textit{homo economicus} agent who discounts the NPV via $(1+r)^{t-1}$, with $r$ being the interest rate. Notice that over 2013-2014, the payback period decreased, although the governmental incentives decreased. This is due to the fall in price per installed \textrm{kW} of a PV system (see Figure \ref{fig:prices}). Also, notice that with the introduction of the \textrm{CE5} the payback period is less volatile. This observation suggests that the social influence between agents plays a crucial role in the diffusion of the PV system in the third phase. From the calibration, we have that $w_{\text{ec}}$ is at least an order of magnitude greater than $w_{\text{soc}}$. This is in line with the results found in \cite{PSM2015TFSS}, where the influence of the communication network is negligible during the first two phases described above. Nonetheless, we point out that the weights coefficients should not be directly compared to each other because of the different formulations in their partial utilities, and their value should be interpreted as their relative importance in the adoption decision process. 
\begin{figure}[h!]
    \centering
        \includegraphics[scale=0.45]{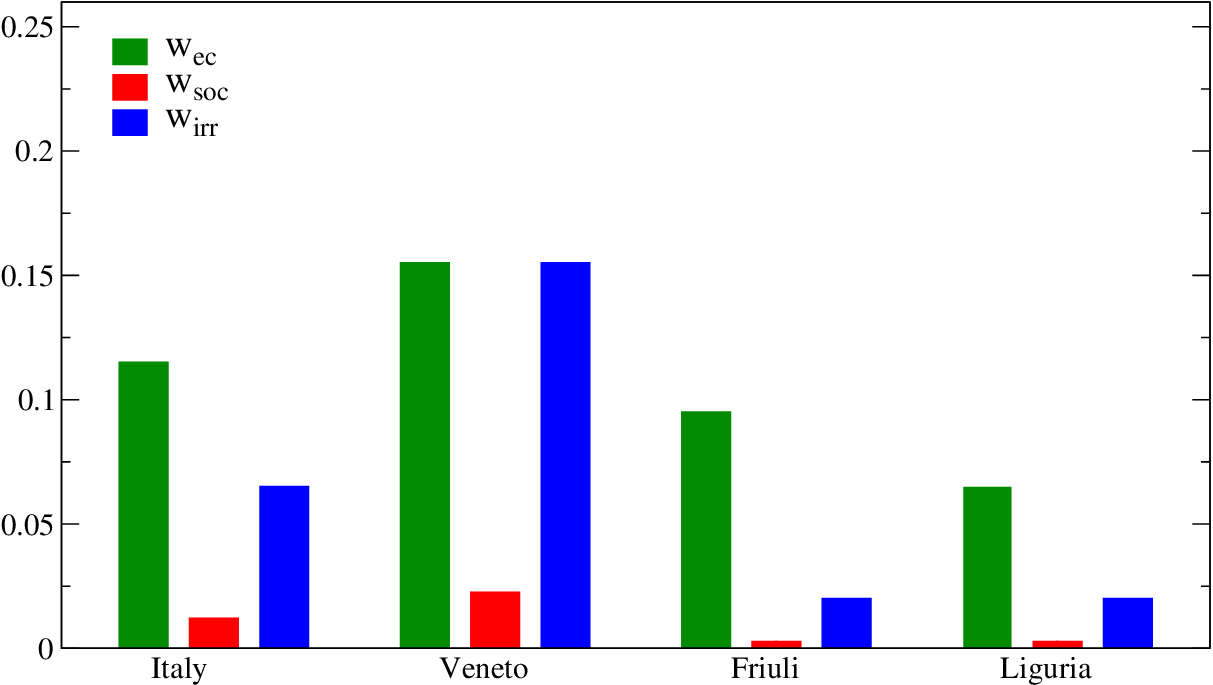}
    \caption{Influence of the weights $w_{\text{ec}}$, $w_{\text{soc}}$ and $w_{\text{irr}}$ obtained by calibrating the model on the $2/3$ of times series of Italy and of the three different regions.}
    \label{fig:histogram}
\end{figure}
\begin{table}
\begin{center}
\begin{tabular}{|l|l|l|l|l|l|l|}
\hline
        & $w_{\text{ec}}$ & $w_{\text{soc}}$ & $w_{\text{irr}}$  \\ \hline
Italy   & $0.11$ & $0.0076$    & $0.06$  \\ \hline
Veneto  & 0.15 & 0.0175  & 0.15   \\ \hline
Friuli  & 0.09 & 0.001  &0.015   \\ \hline
Liguria & 0.06 &  0.001  &0.015  \\ \hline
\end{tabular}
\caption{Optimal value for the fitted weights $w_{\text{ec}}$ and $w_{\text{soc}}$ obtained by calibrating the model on the $2/3$ of times series of Italy and of the three different regions.}
\label{tab:weights}
\end{center}
\end{table}
\begin{figure}
    \centering
    \includegraphics[scale=0.39]{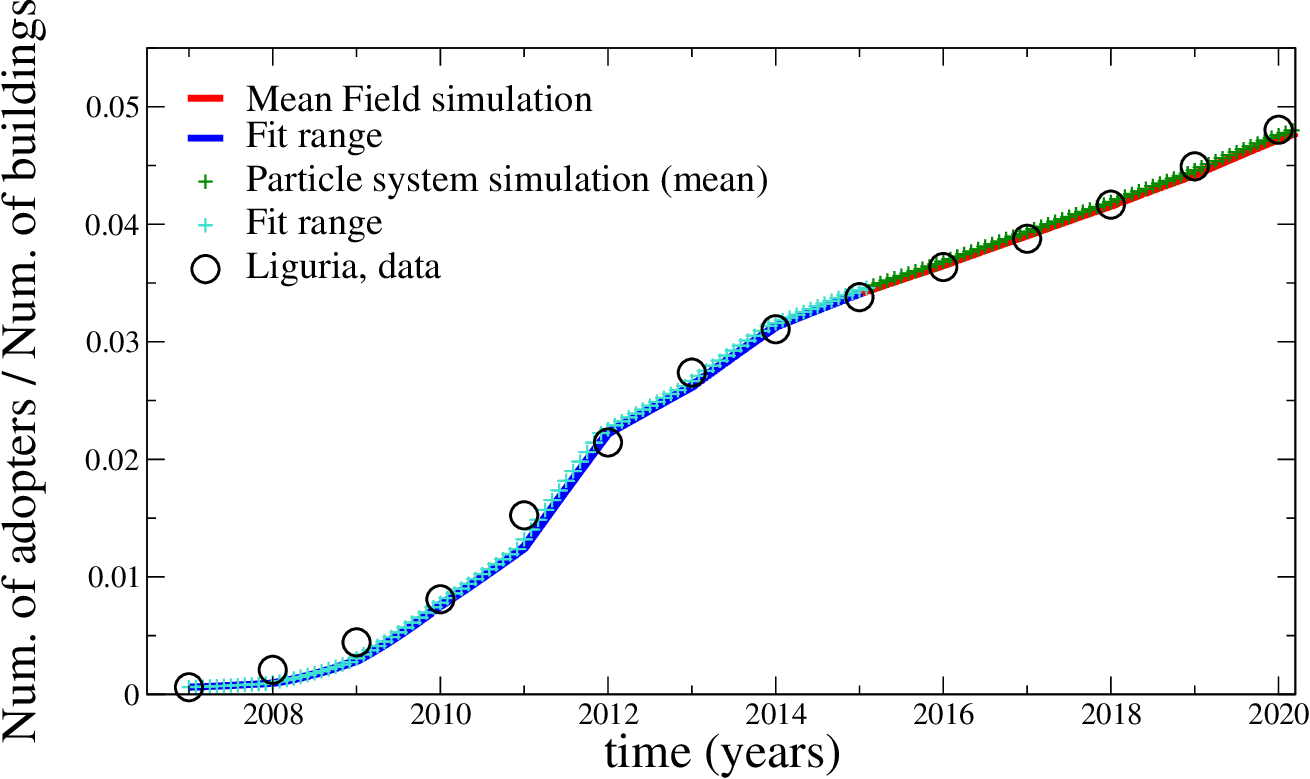}
    \includegraphics[scale=0.39]{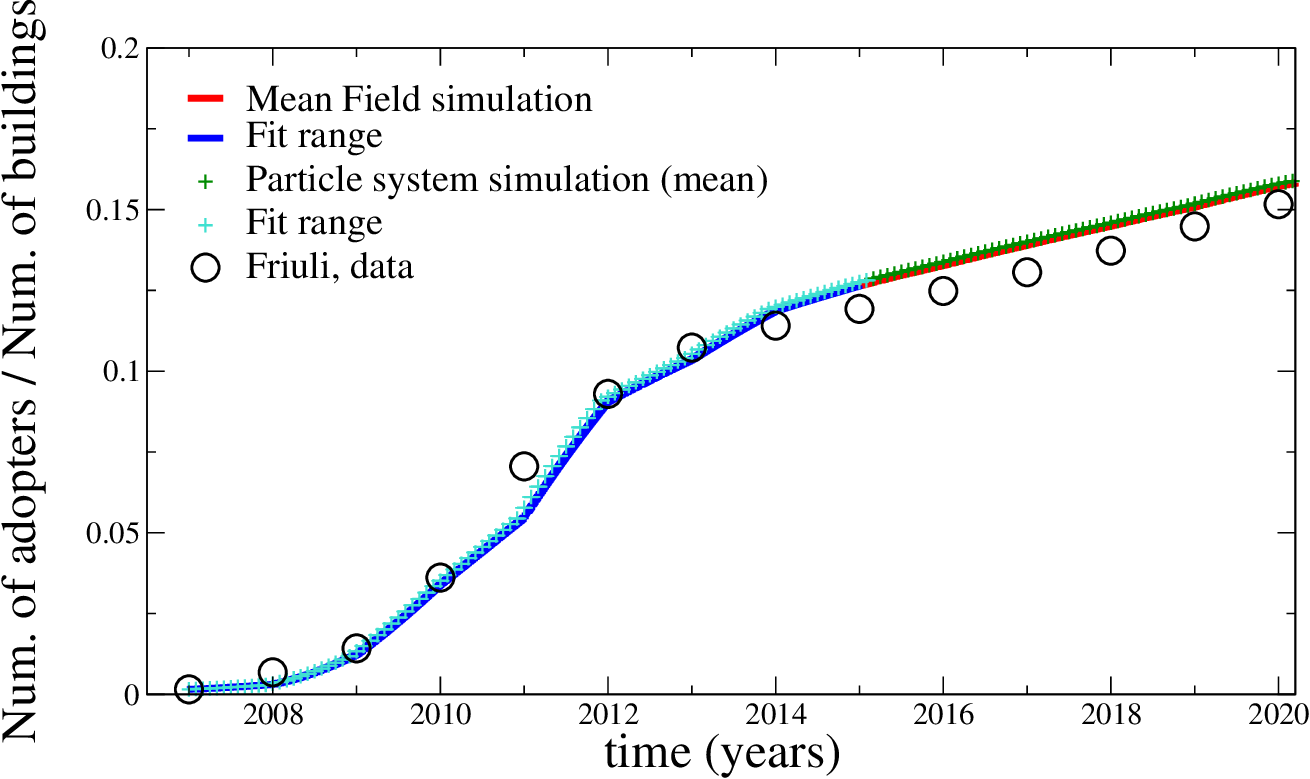}
    \includegraphics[scale=0.39]{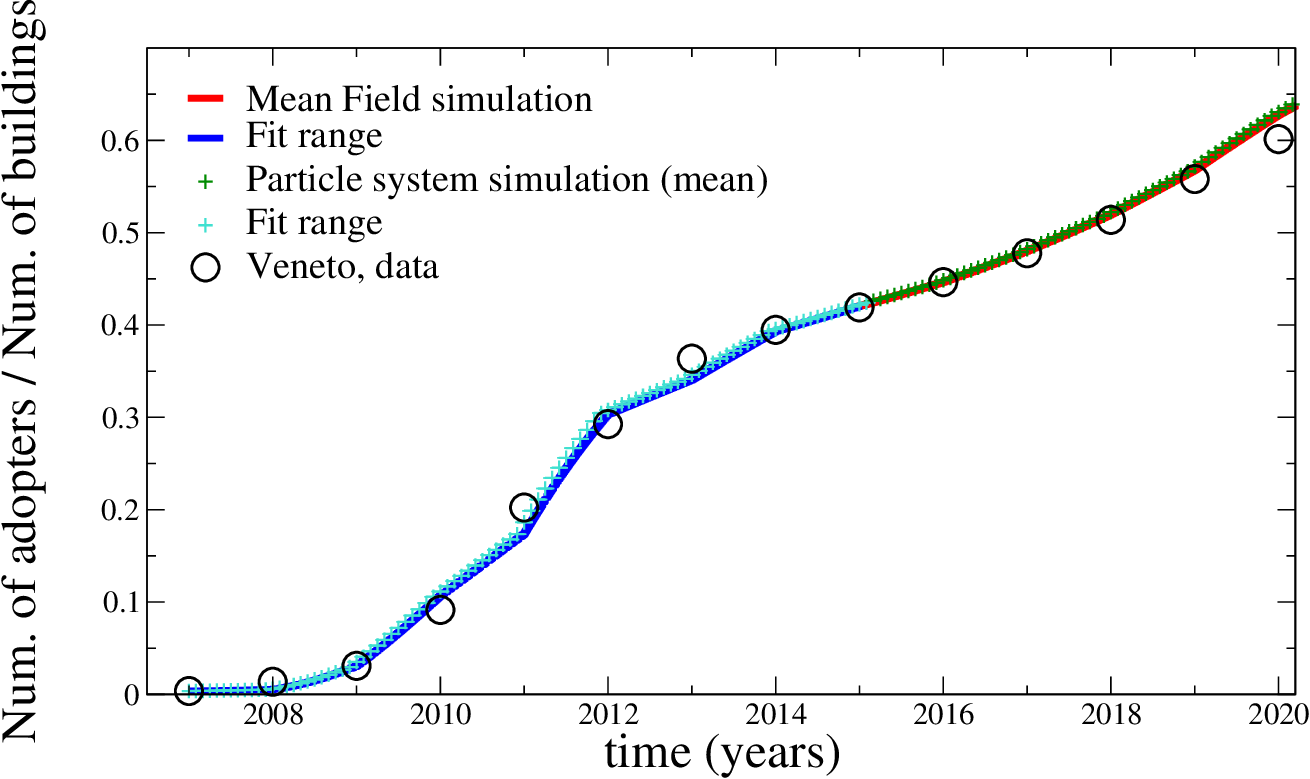}
    \includegraphics[scale=0.39]{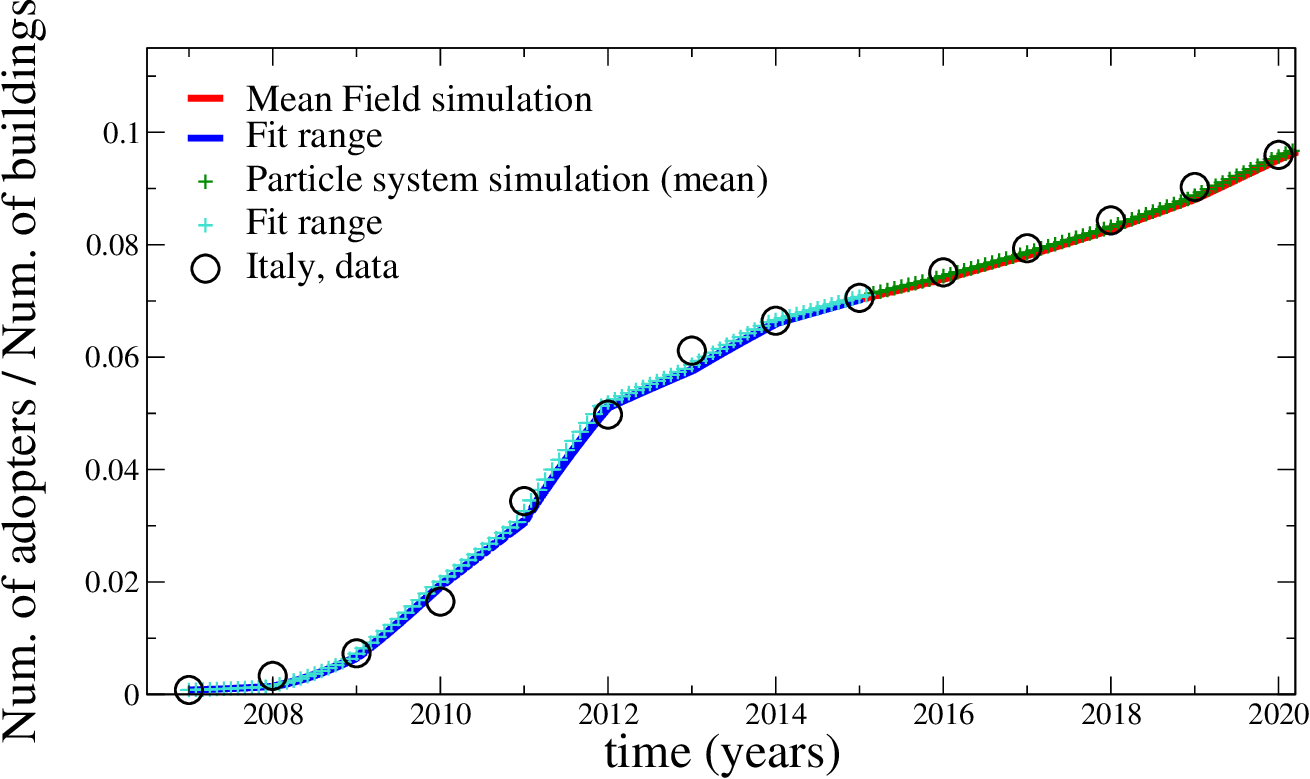}
    \caption{Calibration of the installed PV capacity, 2007--2020 for Liguria, Friuli Venezia Giulia, Veneto, and Italy. \textit{Source\,:\,} Own illustration, based on calibration results.
    The blue line (resp. light blue dots) represents the calibrated mean field model (resp. particle system) applied to the first two-thirds of the data set, while the red line (resp. green dots) shows the mean field model (resp. particle system) run during the remaining period. }
    \label{fig:regions}
\end{figure}
\begin{figure}
    \centering
    \includegraphics[scale=0.45]{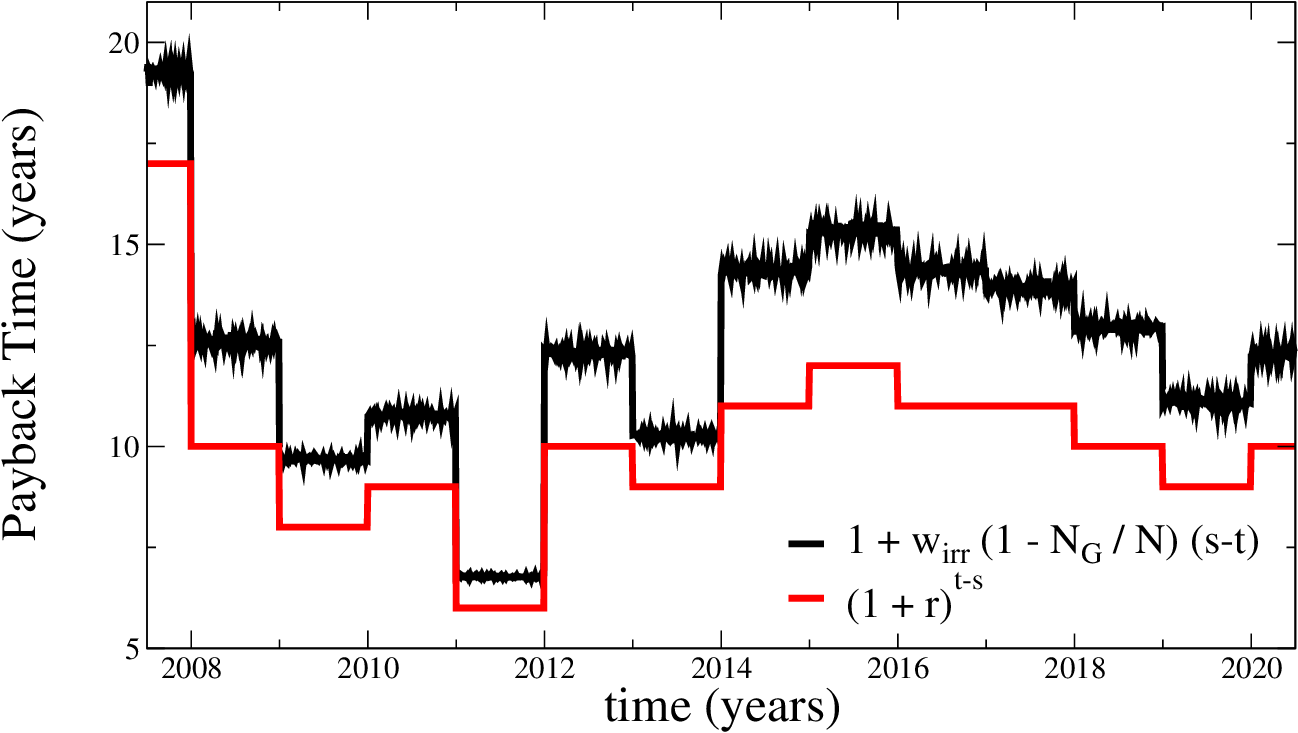}
    \caption{Example of a payback period of a PV system over 2007-2020, with the discount of the so called ``homo economicus'' and our discount where for each time $t$, $W_{\text{irr}}$ is a random variable with a triangular distribution with average $w_{\text{irr}}=0.06$. }.
    \label{fig:pp}
\end{figure}

\clearpage
\subsection{Stability analysis}\label{subsec:simulationsresults_stability}
This subsection discusses the robustness, with respect to the parameters $k_I, k_P$ and $T$, of the results reported in the previous Subsection \ref{subsec:modelcalibration}. Because we do not study the convexity properties of the likelihood of our model, we point out that there may be other combinations for the values of $k_I, k_P$ and $T$ that lead to similar, or even better, fitting results. We will report the stability analysis' results for the Italian photovoltaic market; the results for the single regions are available from the authors upon request. Figure \ref{fig:simulationsresults_stability} summarises the results. In its top panel, we vary $k_I \in [9.5,14]$, and we keep fixed the values of $k_P$ and $T$. In the middle panel of Figure \ref{fig:simulationsresults_stability}, we vary $k_P \in [9.5,14]$, and we keep fixed the values of $k_I$ and $T$. Finally, in the bottom panel, we vary $k_I \in [3,7]$. Overall, the results are satisfactory; however, we should notice that this procedure appears to be more stable when calibration is conducted over a somewhat larger portion of the time series than the one shown in Figure \ref{fig:regions}. Numerically, we verify that it is more stable when applied to $7/10$ of the time series rather than $2/3$. We now make the following important observations. First, we observe that the values of the weights, although different in values, maintain the same relations, i.e., $w_{\text{soc}}\leq w_{\text{irr}}\leq w_{\text{ec}}$. Second, we observe that by increasing the initial number of "Planner" and "Informed", i.e., by increasing the values $k_P$ and $k_I$, we obtain, quite naturally, lower values for $w_{\text{ec}}$ and $w_{\text{soc}}$. Indeed, by increasing $k_P$ and $k_I$, we are increasing the segment of the population that is close to making the transition. Finally, we observe that by increasing the value of $T$, we obtain higher optimal values for $w_{\text{irr}}$. This result is consistent with the model's expectations. In fact, as $T$ increases, the projected economic utility would rise, thereby expanding the pool of individuals close to making the transition. With a fixed $w_{\text{irr}}$, this would result in an unexpected increase in the number of subjects making the transition, making it impossible to achieve a proper fit. More precisely, both $T$ and $w_{\text{irr}}$ are parameters linked to the motivational barriers and are positively correlated; the further into the future an individual projects the event, the more irrational they are.

\begin{figure}[h!]
    \centering
    \includegraphics[scale=0.45]{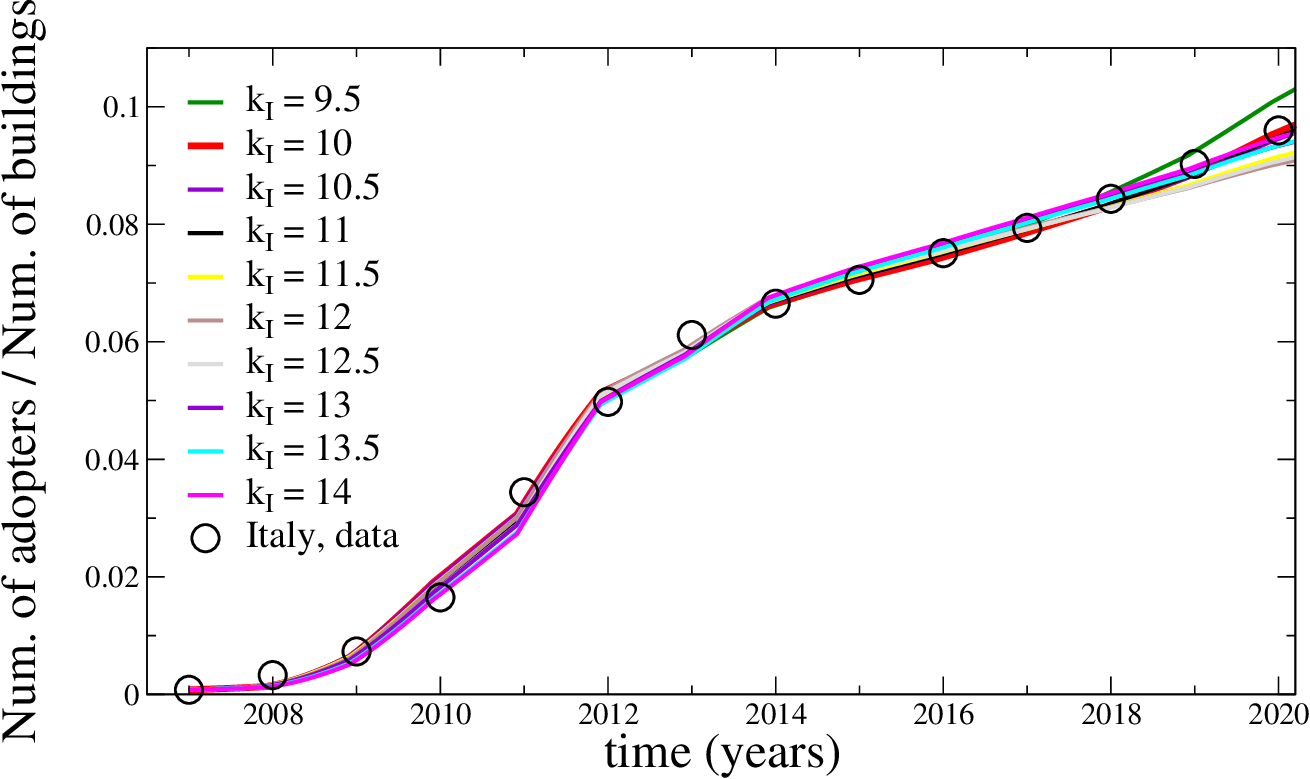}
    \includegraphics[scale=0.45]{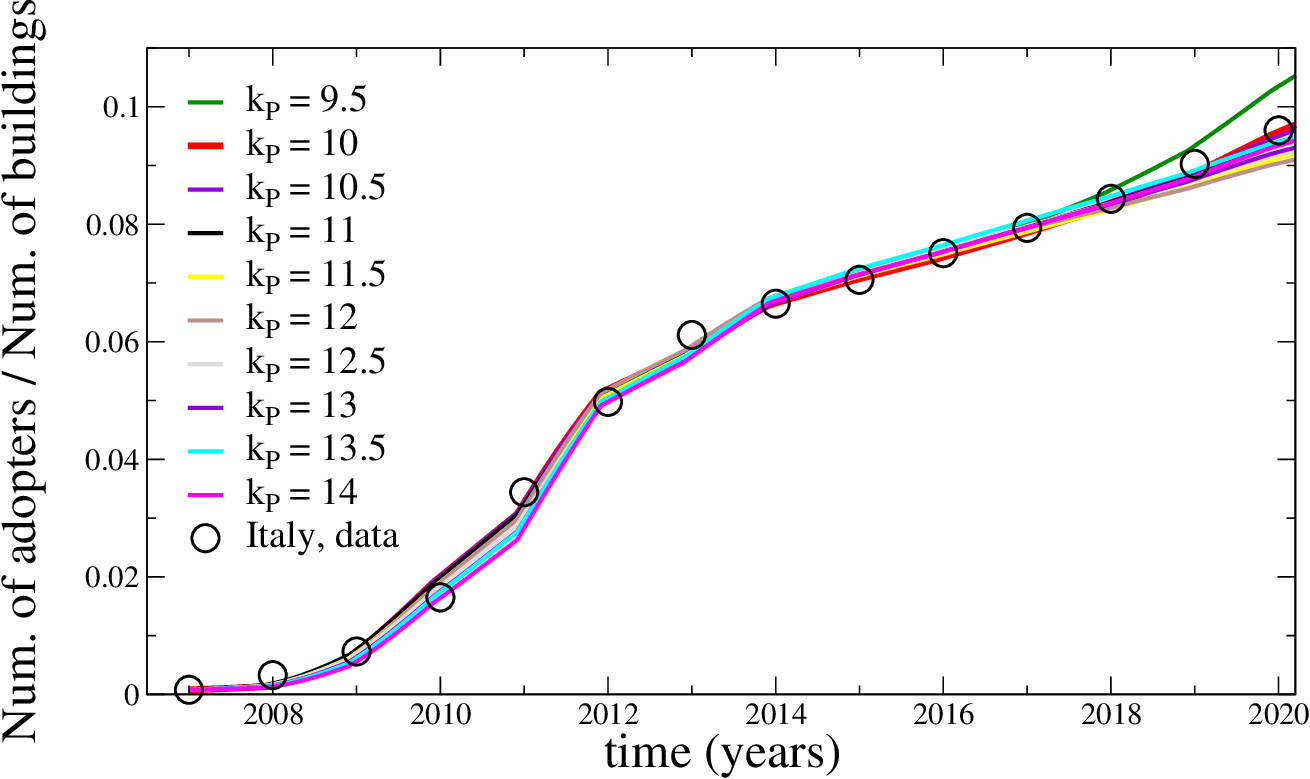}
    \includegraphics[scale=0.45]{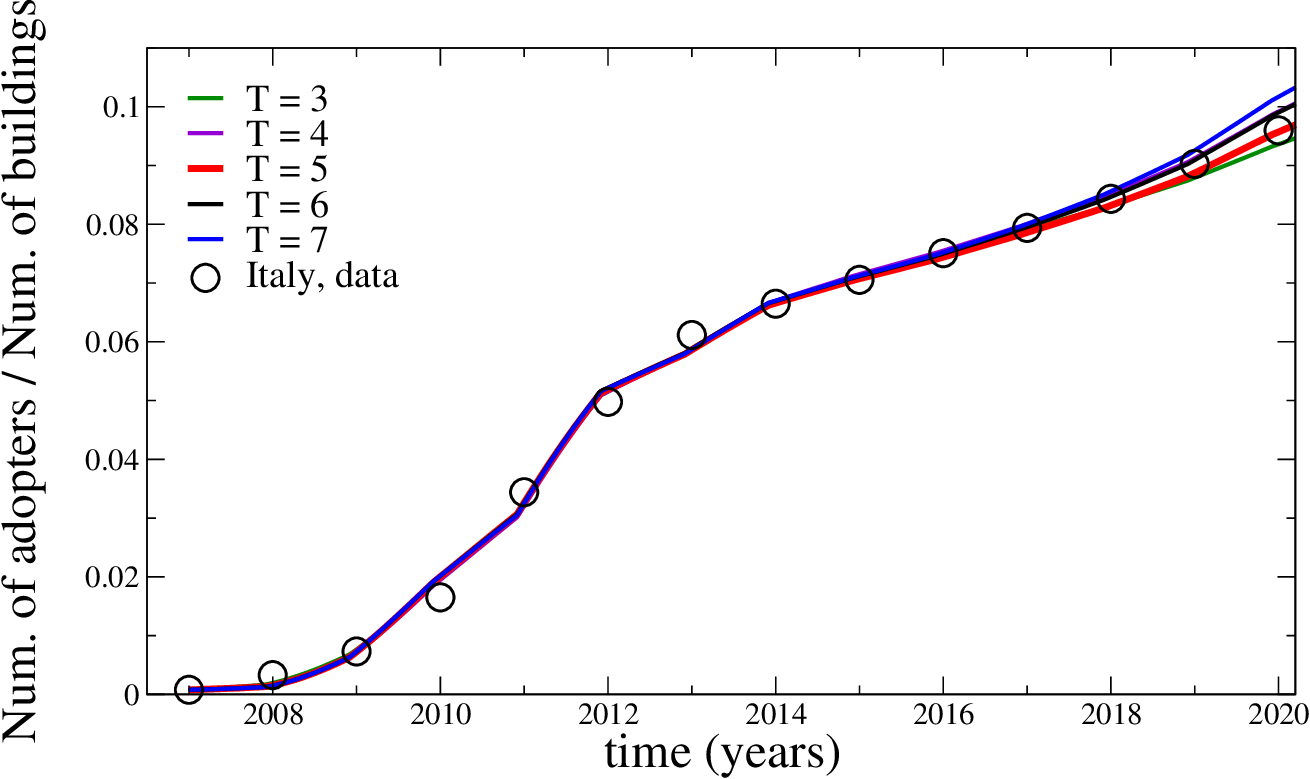}
    \caption{Stability analysis on the initial number of agents in each state.  \textit{From top to bottom\,:\,}  $k_P$ and $T$ fixed;
     $k_I$ and $T$ fixed;
     $k_I$ and $k_P$ fixed.}
\label{fig:simulationsresults_stability}
\end{figure}

\subsection{Sensitivity analysis}\label{subsec:sensitivity}
This section aims to explore the model's sensitivity to the parameter used to do the calibration. The results of the sensitivity analysis is summarized by Figure \ref{fig:sensitivityone}, which displays the sensitivity concerning the weights (description in items (I), (II), and (III) below).
Sensitivity analysis is performed by holding constant the values of the calibrated parameters in the national data adaptation, see the first line of Table \ref{tab:weights} and varying the parameter whose sensitivity is analyzed. In particular:
\begin{itemize}
    \item[(I)] The weight of the payback period $w_{\text{ec}}$ has,  due to the linear formulation of its partial utility, a stronger impact on the diffusion process than that of the other weights. Indeed, our agents in passing from ``Planner" to ``Green" are \textit{homo economicus} agents, which means that if $w_{\text{ec}} \approx 0$, then no transition occurs; see Section \ref{sec:MarkovianModels}. We argue that this causality is not captured by models in which the transition occurs by evaluating a utility function expressed as the sum of weighted partial utilities accounting for different factors (e.g., the environmental benefit of investing in a PV system or the influence of communication with other agents). Indeed, in these models, the transition could happen even if it is not economically convenient. 
    \item[(II)] The weight $w_{\text{soc}}$ plays a very different role than the payback period weight. From Figure \ref{fig:sensitivityone}, \textit{Middle Panel}, we observe that higher is the value of $w_{\text{soc}}$ and closer is our model to a logistic one. On the other hand, if $w_{\text{soc}} \approx 0$, then there is no transition; see the \textit{green} line in the corresponding figure. More precisely, the transition $I \rightarrow PL \rightarrow G$ happens at a much faster rate than the transition $C \rightarrow I$, and, in practice, the Markovian model comes down to a model with states $\{C, PL, G\}$. Because the transition $C \rightarrow PL$ is not more allowed, no further transition is observed. Said differently, the state $I$ is not renovated rapidly enough.
    \item[(III)] The parameter $w_{\text{irr}}$ influences, by construction, only the economic utility. In particular, when $w_{\textit{irr}}$ is one-sixth of the calibrated value, our agents are neither \textit{homo economicus} nor bounded rational (see the discussion at the end of Section \ref{sec:MarkovianModels}, where the same effect is obtained by setting $N_{G}(X_t)=N$), and they perceive a higher economic utility, thus obtaining a similar effect to an increase of $w_{\text{ec}}$; see item (I). On the other hand, 
    an increase of $w_{\text{irr}}$ reflects that our agents may end up trapped in a procrastination loop due to the inter-temporal structure of the green choice. Therefore, the cumulative (normalized) number of adopters is still growing but slowing down.
\end{itemize}

\begin{figure}
    \centering
    \includegraphics[scale=0.45]{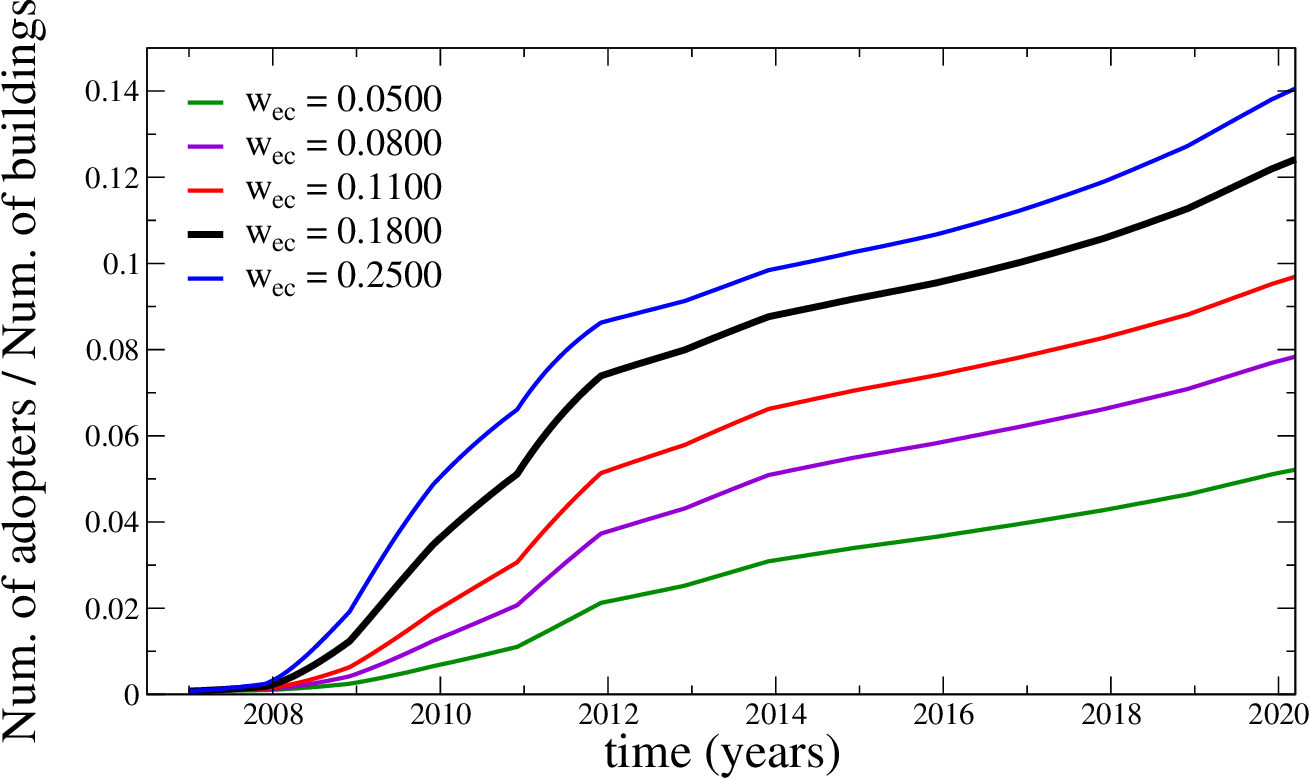}
    \includegraphics[scale=0.45]{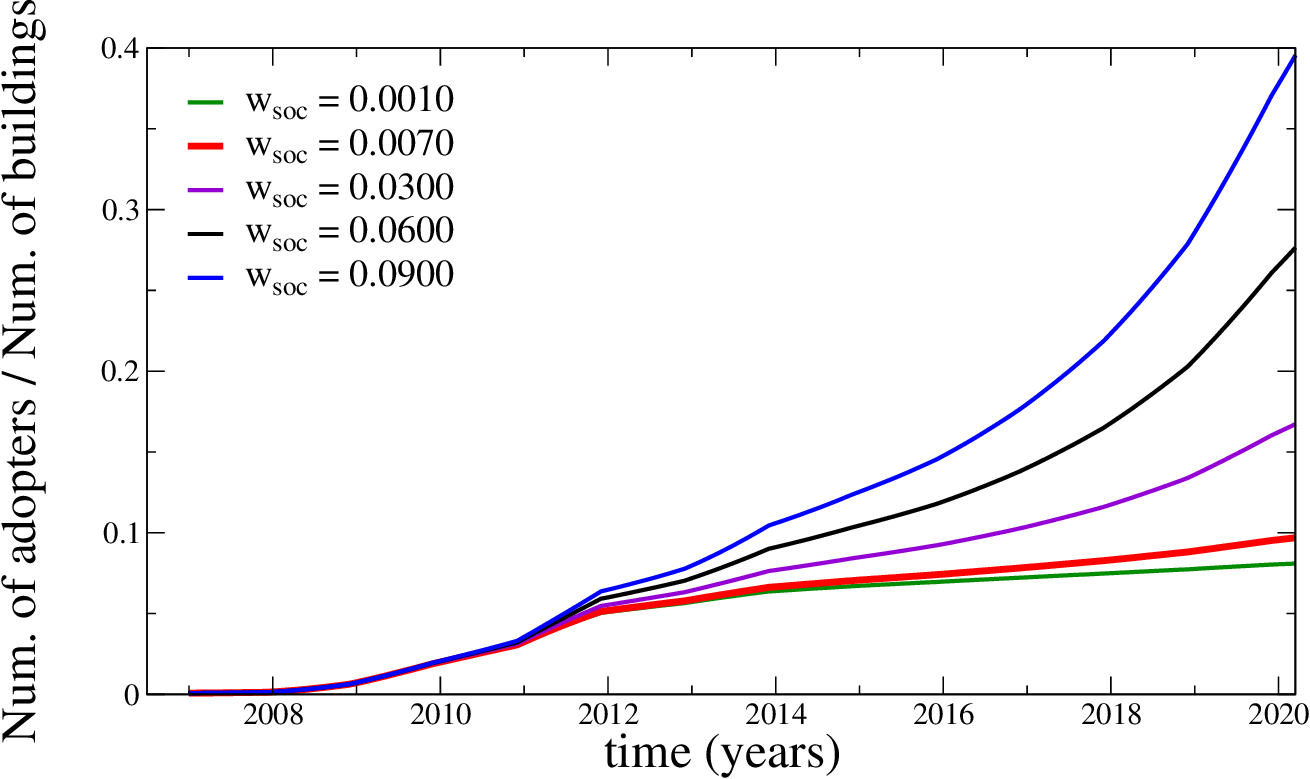}
    \includegraphics[scale=0.45]{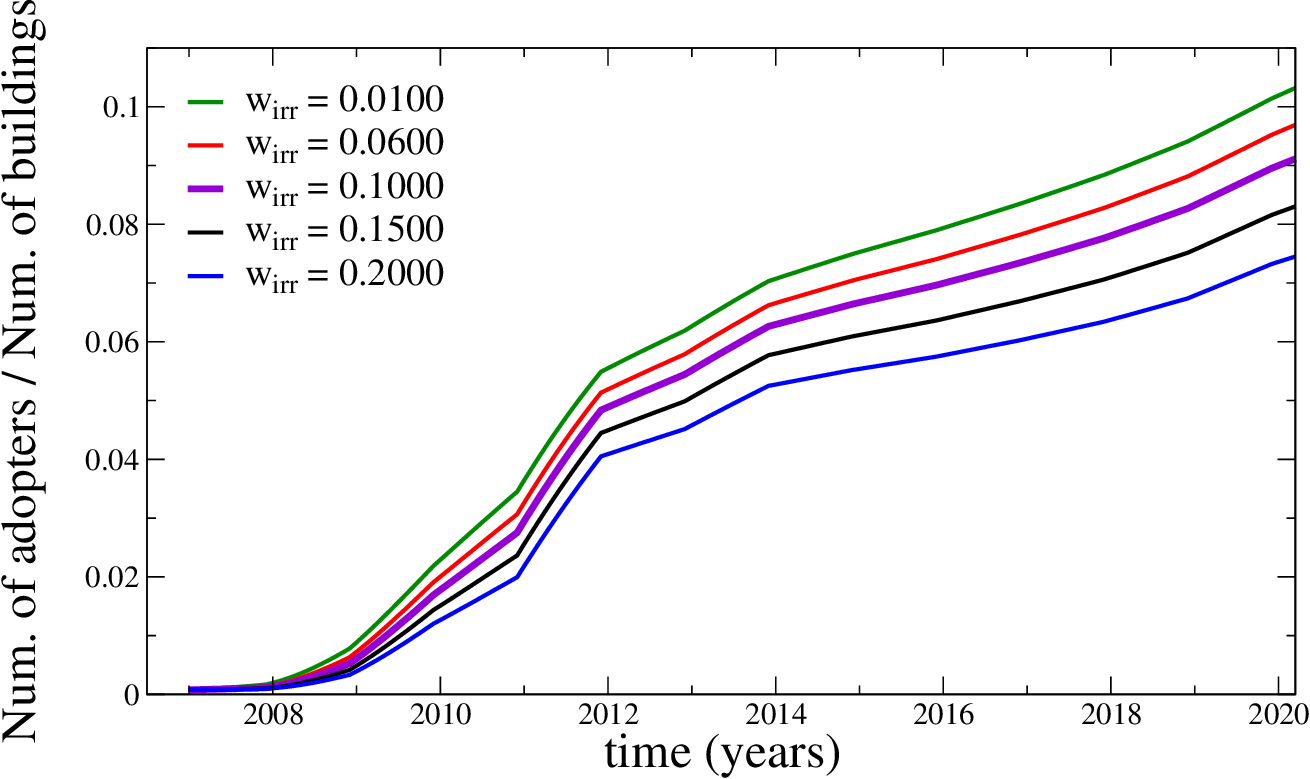}
    \caption{Sensitivity analysis on the weights. \textit{From top to bottom\,:\,} 
    $w_{\text{soc}}$, $w_{\text{irr}}$, $k_P$, $k_I$ and $T$ fixed;
    $w_{\text{ec}}$, $w_{\text{irr}}$, $k_P$, $k_I$ and $T$ fixed;
    $w_{\text{soc}}$, $w_{\text{ec}}$, $k_P$, $k_I$ and $T$ fixed.}
    \label{fig:sensitivityone}
\end{figure}
}

\newpage
\section{Scenario analysis}\label{sec:scenarioanalysys}
We test five different simulation scenarios to consider the sensitivity and validity of the proposed model. The first is a Baseline scenario where we use the set of parameters resulting from the calibration (see Section \ref{sec:numerical_experiments}). Then, we consider a scenario with different PV investment costs (Scenario I), a policy-driven scenario with governmental PV support (Scenario II), a scenario in which a nudging strategy is implemented (Scenario III), and \textcolor{black}{finally}, a scenario in which social interaction is strengthened (Scenario IV). \textcolor{black}{\textcolor{black}{As said}, all five scenarios build on the parametrization obtained from the calibration in Section \ref{sec:numerical_experiments}.}\\
\indent\textcolor{black}{Before examining the results, we will explain the rationale behind the design of the just-mentioned scenarios. Scenario I and Scenario II correspond to Scenario II and Scenario III analyzed in \cite{PSM2015TFSS}, respectively. As in their work, they are justified by the observation that, during the calibration period, the weight associated with economic utility primarily influenced the decision to adopt photovoltaic technology (see Figure \ref{fig:histogram}). There are, however, two key differences with respect to the previous study. First, the calibration period: their calibration period is 2006--2011; so in particular, their parameters are calibrated in a period of moderate-size Feed-in-Tariff subside policy (before 2011) and substantial Feed-in-Tariff subside (around 2011); see  Appendix \ref{app:subsec:evolution_PVs}. Second, they do not consider the impact of the bounded rational behaviors of home owners on the adoption of solar photovoltaic systems.  In Subsection \ref{subsec::scenarioII}, we will see that, as in \cite{PSM2015TFSS}, both governmental incentives and the evolution of the price of the photovoltaic system have a significant influence on the adoption process. However, in our case, Scenario II will obtain a smaller technology adoption with respect to Scenario I, which is consistent with the fact that after 2012 the firm subsidy policy was weaker than the one before 2012. Interestingly, such a discrepancy can be justified in terms of hyperbolic discounting. Moreover, in \cite{PSM2015TFSS} authors do not focus their scenario analysis on the lock-in of environmental behaviors through the following two distinct mechanism: through individual decision making and through social structure. Our modeling framework, instead, allow to test wether governments can use nudges to encourage homeowners to adopt PV. Our ultimate goal is to study the impact of both intertemporal nudges and social contagion policies on PV diffusion, so that they can be taken into account by policymakers when defining energy policy portfolios that lead to rapid decarbonization. This requires analyzing the effects of intertemporal nudges and social contagion not only in absolute terms, but also relative to the effects of traditional climate policies based on prices and incentives. We have therefore developed a four-scenario analysis as the best-fit analytical approach to help policymakers identify the most effective mix of energy policies. Precisely, Scenario III is a self-interested type of nudge. Scenario IV, instead, calls on prosocial, community-oriented motives rather than on self-interest. In Subsection \ref{subsec::scenarioIII}, we will use a framing that induces the individuals who want to install the PVs to agree
on the installation at a certain date and to start paying for it at a specific date in the future. We will see that this self-interested type of nudge increase the pressure to act. In addition, we will see in Scenario IV that an increase in $w_{\text{soc}}$\footnote{An increase in $w_{\text{soc}}$ can be obtained, for instance, by sending to homeowners a personalized letter activating social norms regarding one's community by demonstrating how many of one’s neighbors have already installed PV.} help increase adoption. These findings are important as they provide governments with a concrete additional tools which leverage psychological insights to motivate homeowners to adopt PV. In this regard, we mention the following studies. The recent work of \cite{neumann2023governments}, where authors conduct a preregistered field experiment involving 600 homeowners in Switzerland, testing whether two types of personalized behavioral interventions, one based on prosocial motives and one focusing on self-interest, lead to tangible actions towards PV adoption. The work of \cite{zhang2022agent}, where authors set the following two different scenarios: (1) Network structure; (2) subsidy strategy, to investigate factors of interest on residential PV diffusion in Singapore. Finally, the work of \cite{masini2003forecasting}, where authors focuses on the situation of southern Europe in the early 2000s, in two specific market segments: (i) building-integrated systems for bulk electricity production in small and remote islands provided with a local electricity grid; (ii) building-integrated systems for domestic grid support. This work takes into account indicators such as the PV shipments, the expected evolution of PV module cost, the expected date when PV will become competitive and the net avoided carbon emissions. The conclusions of the study are that the adoption of policies such as inter-temporal nudging and the strengthening of communication networks can contribute significantly to reducing emissions and consequently market inefficiencies. This is important because these policies have much lower implementation costs than traditional policy instruments such as taxes and subsidies. Our findings are similar to \cite{masini2003forecasting}, in the sense that nudging and the facilitation of communication between agents seems to be equally or more effective than the action on the economical benefits.}\\
\indent \textcolor{black}{In addition,} \textcolor{black}{we remind that our Markovian-type framework cannot model unpredictable events beyond what is typically expected of a situation and have potentially severe consequences, i.e., black swans.  For this reason, we have not considered incentive schemes for installing PVs starting in 2023, and calibrated the model until 2020; indeed, during this year, the Italian economy is still on the path to recovery from the COVID-19 pandemic and is affected by the armed conflict in Ukraine.} Likewise, the five scenarios must be contextualized to a standard economic environment.\\
\indent For the reader's convenience, we grouped the figures related to the scenario analysis at the end of the present section, in Subsection \ref{subsec:fig}, in the order that they will be \textcolor{black}{mentioned in the main text}. \textcolor{black}{Besides}, we will report the scenario analyses' results for the Italian photovoltaic market; the results for the single regions are available from the authors upon request. \textcolor{black}{Finally, all scenarios were realized by simulating the mean-field approximation; see Equation \eqref{eq:mf_approximation}.}
\subsection{Baseline scenario}
\begin{figure}[h!]
    \begin{center}
    \includegraphics[scale=0.45]{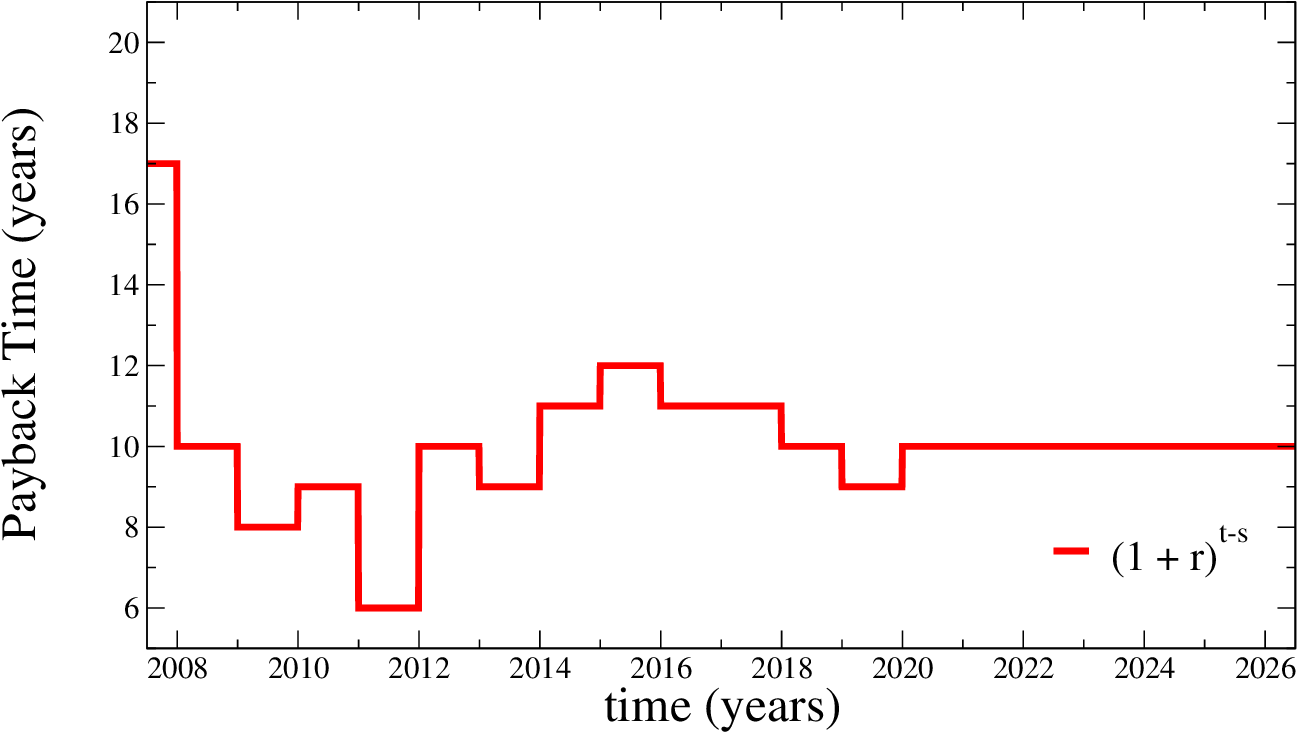}
    \includegraphics[scale=0.45]{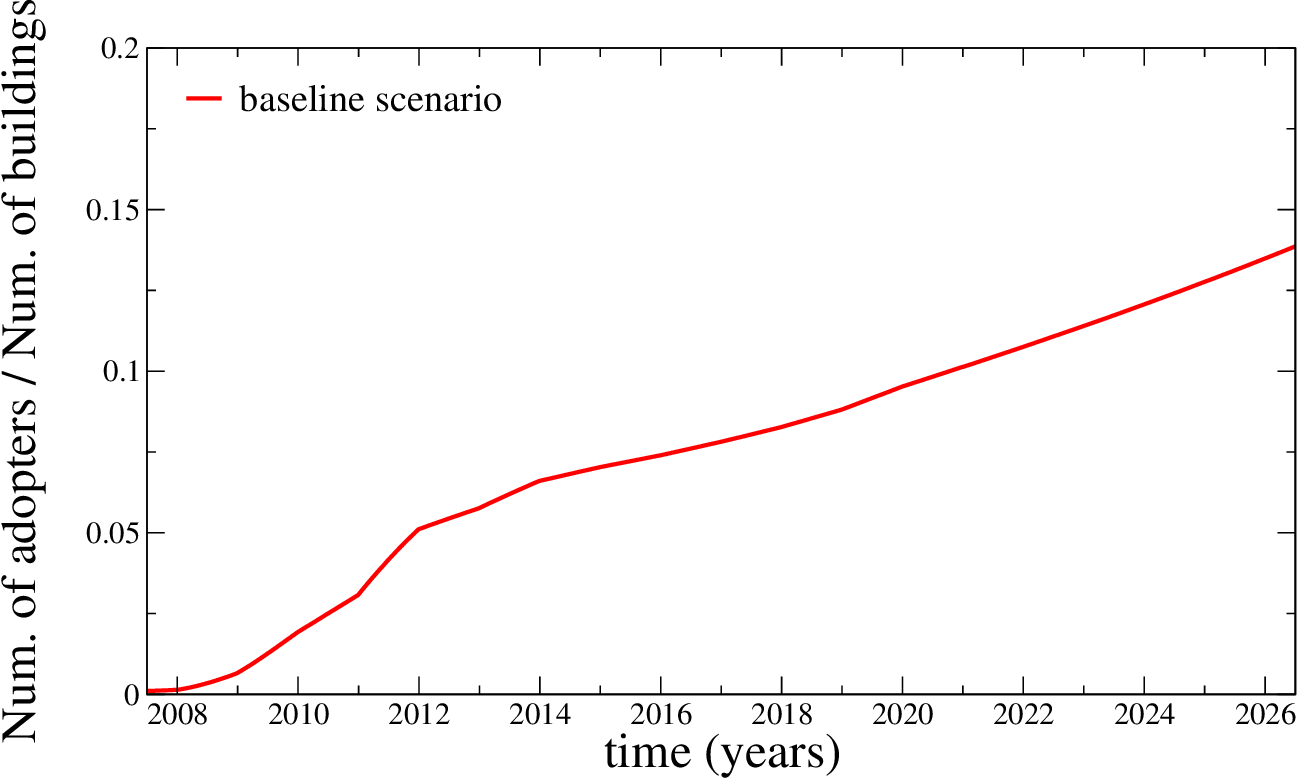}
    \caption{Italy, Baseline scenario. \textit{Top Panel}: payback period of a PV system; \textit{Bottom Panel}: number of installed PV system. \textit{Source\,:\,} Own illustration.}
    \label{fig:baselinescenario}
    \end{center}    
\end{figure}

The Baseline scenario considers no further development of the Italian PV market throughout the simulation period \textit{ex-post} the calibration (i.e., 2020-2026) and serves as a comparison with the other scenarios. In accordance, the payback period remains constant and equals its 2020 value; see Figure \ref{fig:baselinescenario}, \textit{Top Panel}. Understanding the Baseline scenario can help us understand the decision-making process for the type of agents described in our model. The number of adopters will increase by $50\%$ from 2020 to 2025. As explained in Section \ref{sec:numerical_experiments}, the influence of the network, social utility, and communication, in general, is significant in what we have denominated the ``third phase", in which the growth begins to slow mainly because of the elimination of the incentives. Therefore, in the Baseline scenario, the observed exponential increase in the third phase is primarily due to the communication network; see Figure \ref{fig:baselinescenario}, \textit{Bottom Panel}. 

\subsection{Scenario I}
\begin{figure}[h!]
    \begin{center}
    \includegraphics[scale=0.45]{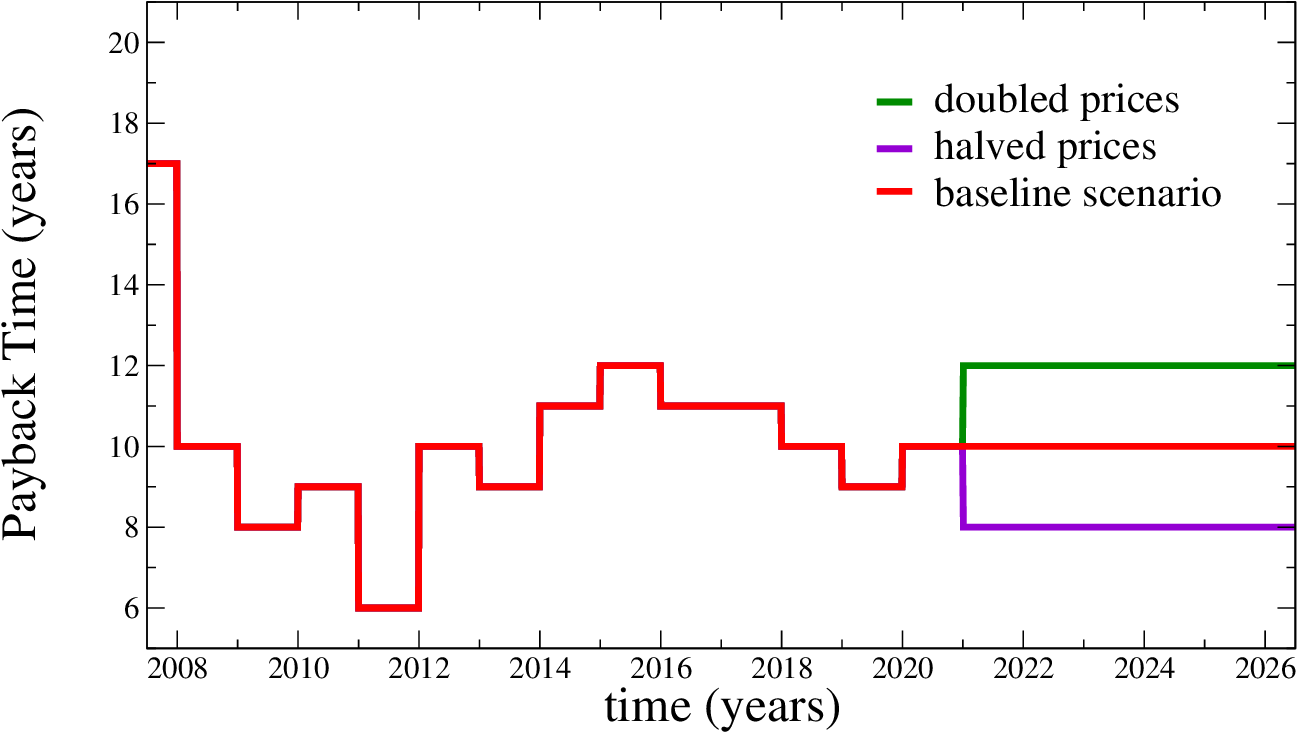}
    \includegraphics[scale=0.45]{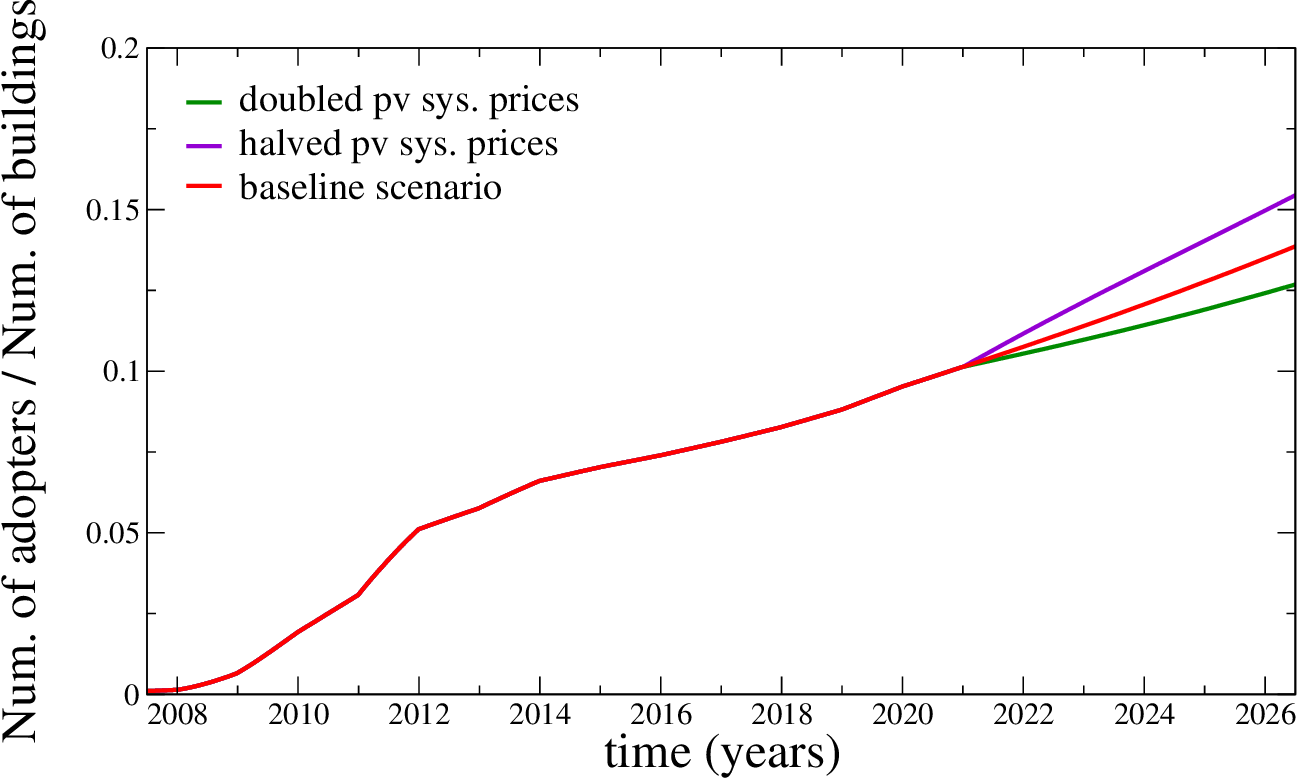}
    \caption{Italy, scenario I. \textit{Top Panel}: payback period of a PV system; \textit{Bottom Panel}: number of installed PV system. \textit{Source\,:\,} Own illustration}
    \label{fig:fut_prices}
    \end{center}
\end{figure}

The first scenario simulates two alternatives for the development of PV system prices. The two alternatives are based on an optimistic and a pessimistic outlook regarding future PV market development from the consumers' perspective. We obtain the ``high" PV system price alternative by increasing the PV system prices by $50\%$. On the other hand, we get the ``low" one by decreasing the PV system prices by $50\%$. Again, our model lacks realism over 2020-2023, so we will not compare numerical simulations with actual data. Figure \ref{fig:fut_prices} collects the results. There is a clear difference relative to the Baseline scenario. A reduction in the investment costs leads to an increase in the total number of adopters of $17\%$ ca  w.r.t the reference case at the end of the simulation period. In contrast, an increase in the investment cost slows the deployment process by $9 \%$ ca w.r.t. the Baseline scenario. This result is not surprising since the economic profitability of the investment is one of the most influential criteria in the adoption decision; the criteria enables the transition from ``Planner" to ``Green". As a result, an increase (resp. a decrease) in the investment costs leads to a reduction (resp. an increase) in the payback period, in this case of two years; see Figure \ref{fig:fut_prices}, \textit{Top Panel}. Another parameter that most influences the investment's economic profitability is the governmental support scheme, which characterizes Scenario II in the following subsection.  

\begin{remark}
We observe that introducing a carbon tax will produce an effect similar to that induced by the increase in initial investment. The introduction of a carbon tax would imply an increase in both prices $p_{purchase}$, $p_{sell}$, and thus $R_{save}$. This fact would make installing photovoltaics even more profitable and increase the number of adopters. 
\end{remark}

\subsection{Scenario II}\label{subsec::scenarioII}
\begin{figure}[h!]
  \begin{center}
    \includegraphics[scale=0.45]{FIG/npv_prices.eps}
    \includegraphics[scale=0.45]{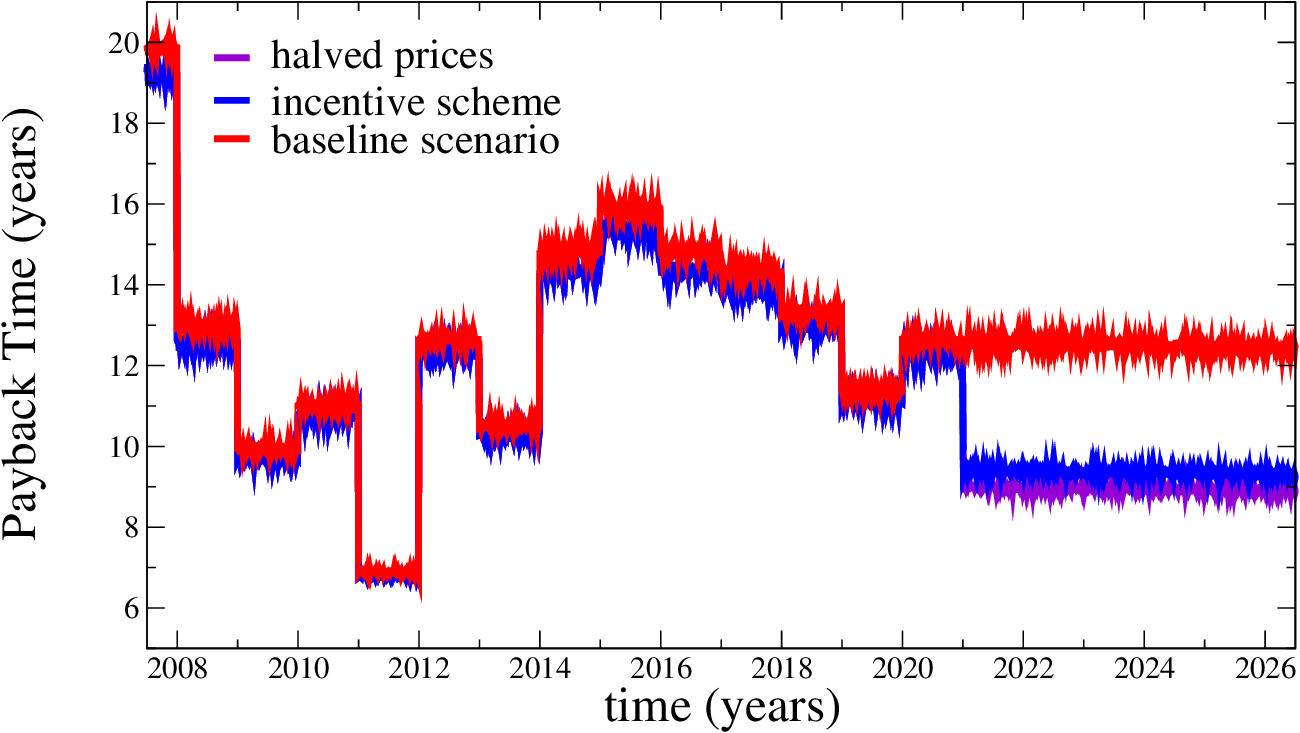}
    \includegraphics[scale=0.45]{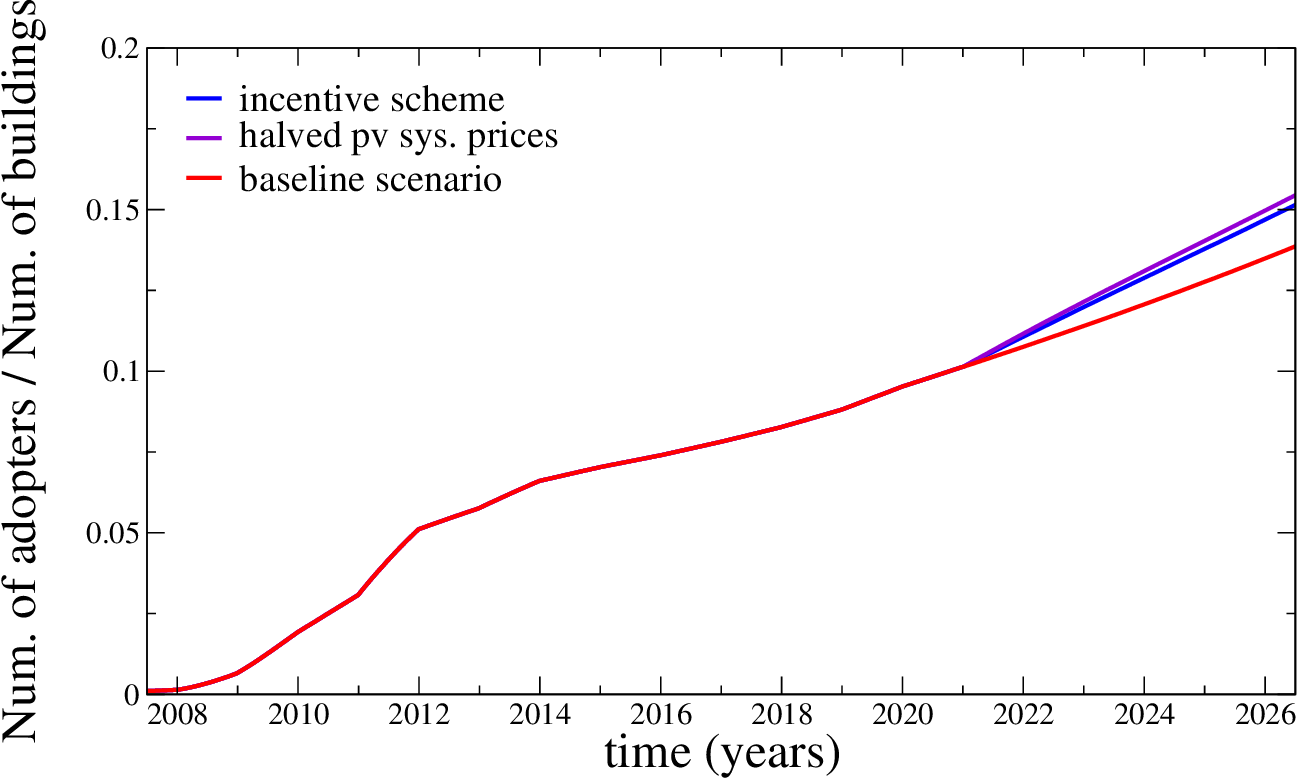}
    \caption{Italy, Scenario II. \textit{Top Panel}: payback time of a PV system. \textit{Middle Panel}: payback time discounted with the hyperbolic discount. \textit{Bottom Panel}: number of installed PV system. \textit{Source\,:\,} Own illustration.}
    \label{fig:incentives}
   \end{center}
\end{figure}

In this scenario, we implement a governmental incentive scheme. Again, we use the Baseline scenario as a reference for comparison. Changes to the support scheme occur from 2020 onward and produce the same effect on the payback period as halving prices; see Figure \ref{fig:incentives}. More precisely, we maintain the same 2020 tax deductions, and we add a bonus equal to $E_{\text{PV}} \times 0.315$, where the quantity $E_{\text{PV}}$ is defined in Appendix \ref{app:investment_costs_and_cash_flows}. The numerical result again shows a difference from the baseline scenario. Compared to the latter, we observe an increase of $10\%$ in the number of PV installations at the end of the simulation period. One observation is in order. Although Scenario I and Scenario II produce a similar change in the payback period, the former leads to a slighter higher number of adopters than Scenario II; see Figure \ref{fig:incentives}, \textit{Bottom Panel}. The hyperbolic discounting explains this discrepancy; see Figure \ref{fig:incentives}, \textit{Middle Panel}. The hyperbolic discounting of future utility can be seen as a temporary weakening of individual rationality induced by the approaching possibility of gains in the present. Hyperbolic discounting differs from exponential discounting of future utility, reflecting rational motives, such as considering the opportunity cost of capital (\cite{A2012TAD, ANG2012JVE, BMDSF2021}). When looking at future choices, most people apply an exponential discount rate, which remains constant over time. For example, subject \textrm{A} might prefer to cash in EUR 1,000 in 2030 rather than wait and cash in EUR 1,100 in 2035, but \textrm{A} might agree to postpone the cash-in if they got EUR 1,200 in 2035. In other words, the opportunity cost of tying up a capital of EUR 1,000 is for \textrm{A} between EUR 100 and 200. \textrm{A} applies a discount rate to the gain only for the time position the gain occupies. Economic theory postulates that if \textrm{A} prefers EUR 1,200 in 2030 to EUR 1,000 in 2035, they must also prefer EUR 1,200 in 2045 to EUR 1,000 in 2040. And for most people, this is indeed the case. However, things change when the choices are not about future investments for even more future earnings but about present investments for future earnings. When the possibility to cash in the present approaches, the individual tends to apply a higher discount rate of future utility than they would apply for future investment choices with the same time distance to earnings. For example, \textrm{A} might be induced to prefer EUR 1,000 today rather than EUR 1,200 in 5 years, even though when faced with a choice between future investments for future earnings, they find it rational to wait five years for a 20$\%$ gain on EUR 1,000. The reasons for this temporary preference for smaller gains in the present are to be found in simple and irrational temporal myopia (\cite{ANG2012JVE}).

\subsection{Scenario III}\label{subsec::scenarioIII}
The third scenario involves implementing a policy to nudge people to transition to solar PV installation by acting on their psyche; we will refer to this policy as \textit{nudging}. Installing PVs is an inter-temporal choice, and the distance between the time of the investment and the time of future earnings is one of the triggers of the procrastination phenomena. Therefore, in the present scenario, we induce a distance reduction between the time of investment and that of future earnings by proposing to the individuals who want to install the PVs to agree on the installation at a certain date and to start paying for it at a specific date in the future.\\
\indent We implement the nudging policy in the following way. We assume that starting from the year 2020, agents pass from $PL$ to $G$ in a time interval $\Delta t \rightarrow 0$ according to the following probability: 
\begin{equation*}
\begin{split}
    &\text{Prob}(X_{t+\Delta t}^{i} = G | X_{t}^{i} = PL) :=  w_{\text{ec}}^{i} \cdot U^{nudg}_{\text{ec}}\left(w_{\text{irr}}^{i}, \frac{N_{G}(X_t)}{N}\right),
\end{split}
\end{equation*}
where the NPV in economic utility $U_{\text{ec}}^{nudg}$ is renewed by shifting the initial cost of the investment into the future time $t+T^*$, i.e., the NPV becomes:
\begin{equation*}
    \text{NPV}(t,n_G(t), \tau) =- \frac{I_{\text{econ}}(t)}{1+g^{i}(t, t+T^*, n_G(t))} + \sum_{s = t+1}^{\tau} \frac{R(s-t)}{1 + g^{i}(t, s, n_G(t)))}.
\end{equation*}
The transition from the state $I$ to the state $PL$ is modified accordingly, through the evaluation of 
\begin{equation*}
    \text{NPV}(t, t+T, n_{G}(t), \tau) = - \frac{I_{\text{econ}}(t+T)}{1+g^{i}(t, t+T+T^*, n_G(t))} + \sum_{s=t+T}^{\tau}\frac{R(s-(t+T-1))}{1+g^{i}(t,s,n_G(t))}.
\end{equation*}

\begin{figure}[h!]
    \begin{center}
    \includegraphics[scale=0.45]{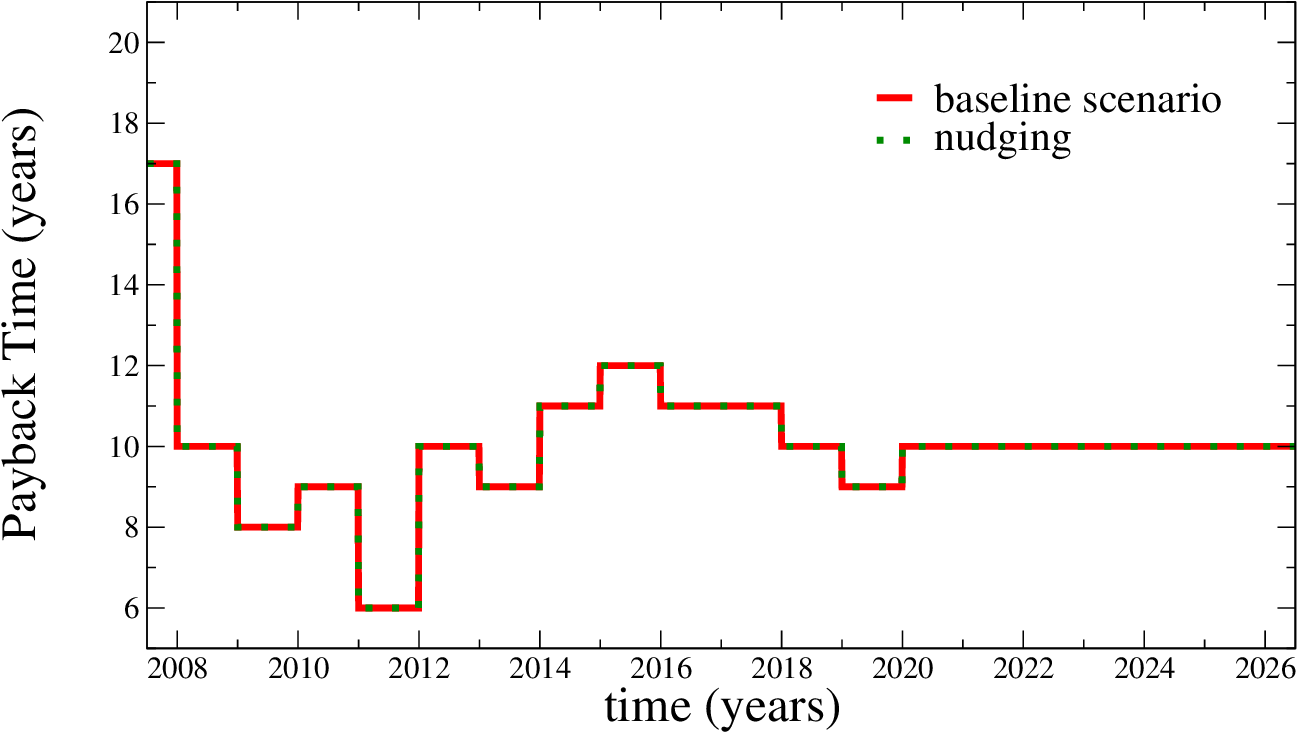}
    \includegraphics[scale=0.45]{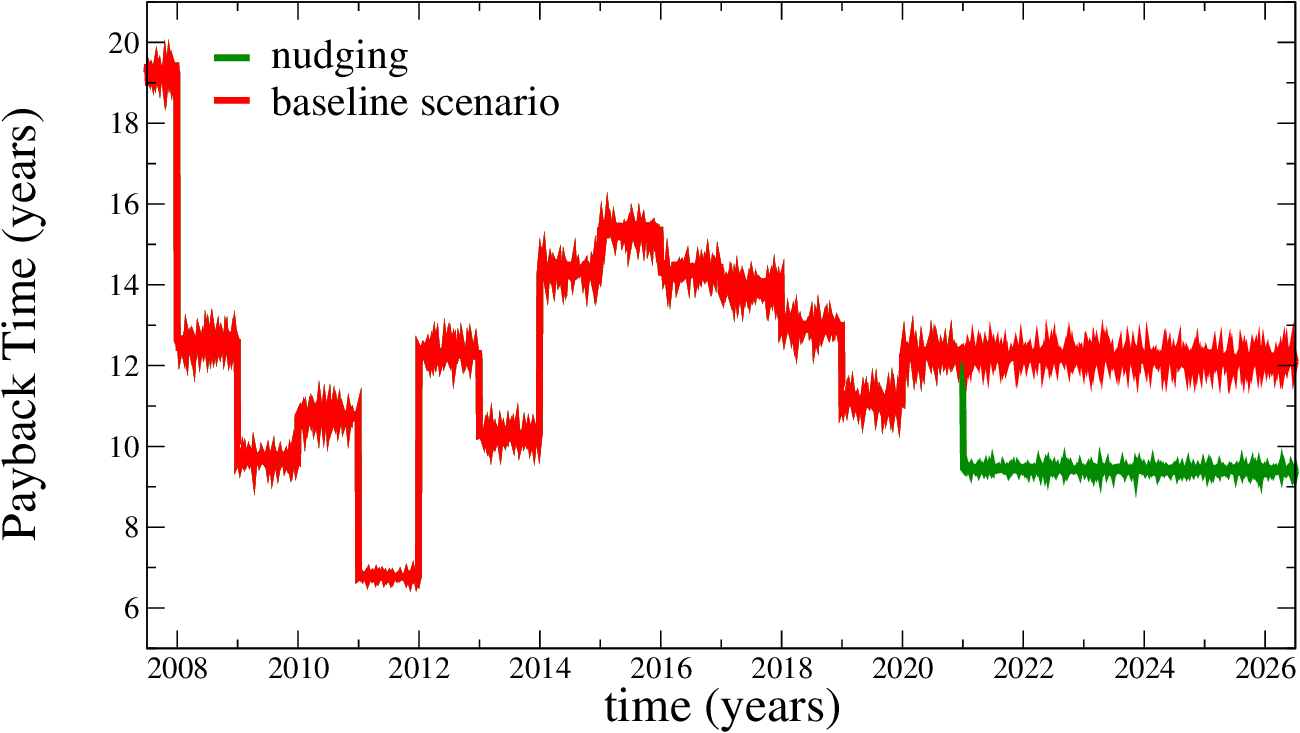}
    \includegraphics[scale=0.45]{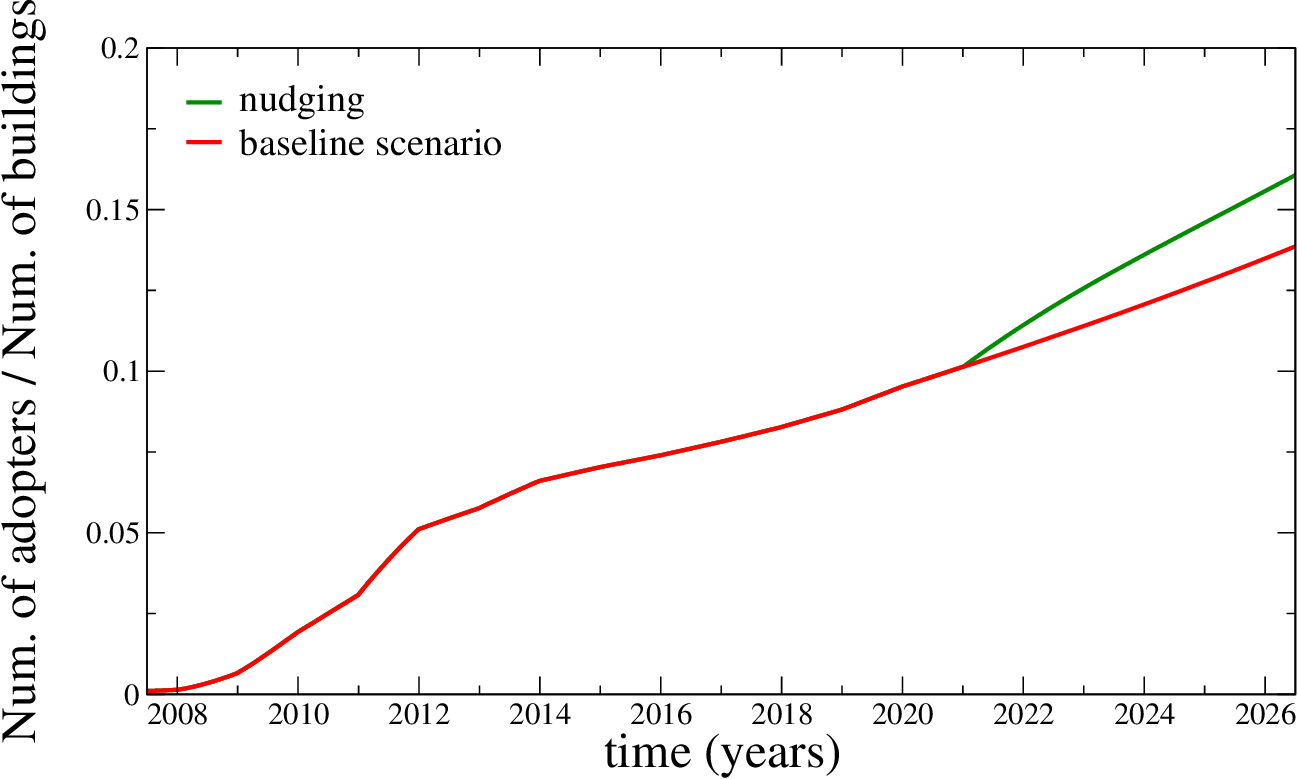}
    \caption{Italy, Scenario III. \textit{Top Panel}: payback period of a PV system; \textit{Middle Panel}: payback period of a PV system;
    \textit{Bottom Panel}: number of installed PV systems. \textit{Source\,:\,} Own illustration.}
    \label{fig:fut_nudg}
    \end{center}
\end{figure}

\noindent Figure \ref{fig:fut_nudg} displays the results when a nudging policy with $T^{*}=1.5$ years is implemented. The results indicate clear differences relative to the Baseline scenario. Nudging leads to an increment of $24.33\%$ in the total number of adopters w.r.t the Baseline scenario at the end of the simulation period. Interestingly, while nudging does not affect the payback period of a \textit{homo economicus} agent (Figure \ref{fig:fut_nudg}, \textit{Top Panel}), it has an effect on agents' payback period that is characterized by bounded rationality because of the presence of the hyperbolic discount (Figure \ref{fig:fut_nudg}, \textit{Middle Panel}). In particular, by postponing the start of the investment, we mitigate the irrational behaviour of the agent linked to procrastination.

\subsection{Scenario IV}
\begin{figure}[h!]
    \centering
    \includegraphics[scale=0.45]{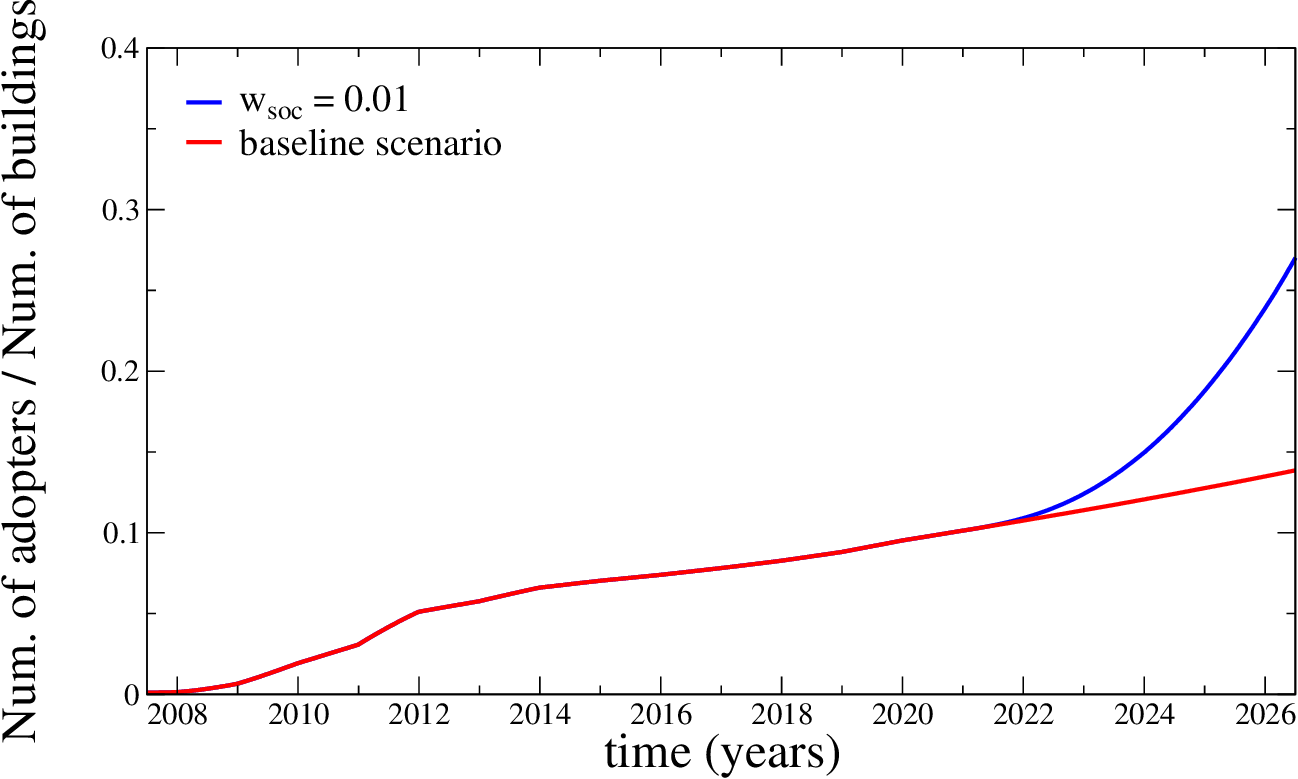}
    \caption{Italy, Scenario IV: number of installed PV systems. \textit{Source\,:\,} Own illustration.}
    \label{fig:future}
\end{figure}

In this last scenario, we propose strengthening the agent's communication network \textcolor{black}{by acting on the value of the weight  $w_{\text{soc}}$}. Figure \ref{fig:future} shows the results if we consider a value of $w_{\text{soc}}$ ten times greater than the calibrated value $w_{\text{soc}} = 0.01$. We observe an increase of $93\%$ in the number of PV by the end of 2026. Moreover, the type of growth makes this scenario different from those previously proposed. Indeed, the increase is exponential in this case, whereas the growth was linear in previous cases. This latter fact can be explained by observing the role of $w_{\text{soc}}$ in the so-called mean-field approximation of the model; see Equations \eqref{eq:mf_approximation}. In particular, the density of the ``Green" $n_G$ increases with the density of ``Informed" agents $n_I$, whose density increases, modulated by $w_{\text{soc}}$ with $n_G$ itself. This fact means that $n_G$ increases exponentially with a rate proportional to $w_{\text{soc}}$. \textcolor{black}{The high sensitivity of the model on the  parameter $w_{\text{soc}}$ can be explained by making the following observation. An analogy can be drawn between epidemiological models and opinion diffusion models. This type of parallelism is quite common; see, for example, the review of \cite{pastor2015epidemic}. The parameter $w_{\text{soc}}$ plays a role similar to that of the contagion parameter (known as $\beta$ in the classic SIR model). Both parameters ($w_{\text{soc}}$ and $\beta$) represent exponential growth rates; the higher their values, the more rapid and pronounced the exponential growth.} In any case, the growth observed in this scenario may be slightly overestimated. This fact could be explained by modelling assumptions. As we have pointed out several times, agents interact in a mean-field way, which might have triggered an overestimation of the effect of social interaction. A more realistic way to describe the interaction between individuals is to consider a network-like structure; we propose exploring this in future work.
\newpage
\subsection{Scenario analysis: Figures}\label{subsec:fig}

\clearpage
\section{Ethical and policy implications}\label{sec::ethical_and_policy_implications}
\textcolor{black}{This section discusses ethical and policy implications that can be inferred from our model.}\\
\indent \textcolor{black}{Before exploring the latter in depth, we would like to outline our theoretical contributions. In this paper, we introduce a robust grounded theory and a Markovian agent-based model tailored for boundedly rational homo-economicus agents. These agents engage in comparable interactions both amongst themselves and with their environment, governed by shared decision-making rules. Our model stands out for its simplicity, characterized by a limited set of parameters. A pivotal aspect of our approach is its capacity to provide a sequential narrative that elucidates the process by which an agent embraces pro-environmental behavior (PV). This framework not only deepens our understanding of the psychological dynamics surrounding the theme of the green energy transition (GET) but also aligns closely with the actual behaviors reported by adopters in survey responses, as evidenced in \cite{KVM2017}. By bridging theory with observed behaviors, we reinforce the validity of our model and its implications for understanding and promoting sustainable choices.} In particular, \textcolor{black}{we focus on two aspects. First, the mathematical framework proposed in Section \ref{sec:MarkovianModels} helps to strengthen the empirical evidence that the intertemporal decisions that are crucial for environmental protection are significantly influenced by cognitive biases such as hyperbolic discounting of individual future utility. This has implications for how governments should think about climate policy portfolios. Second, the results in Section \ref{sec:numerical_experiments} provide empirical evidence that individual consumption choices should be evaluated not only in terms of the emissions they reduce comparatively, but also in terms of the social impacts they have. This may lead to a reformulation of how the problem of causal inefficiency is posed in individual climate ethics.\\
\indent The classical policy approach to climate change relies mainly on economic rationality. The basic assumption is that individuals emit because it is convenient for them to do so, in terms of cost, time, comfort, etc. In order for them to emit less, the government must therefore give them rational incentives to choose ``green" consumption, i.e. based on renewable energy and resources, over ``brown" consumption, i.e. based on fossil fuels (e.g., \cite{baranzini2017, blanchard2023, sterner2024}). Public subsidies are an example of a positive incentive, as they aim to reduce the cost of green consumption for individuals. They are essential where the research and development of certain green technologies is not yet sufficiently profitable to attract sufficient private capital. Carbon pricing, on the other hand, is an example of a negative incentive. It aims to increase the cost of brown consumption in a more or less linear way in order to achieve a socially efficient allocation of the costs of climate change mitigation.\\
\indent This study provides empirical support for the claim that the rational incentives approach to climate policy is at least incomplete. Policy makers should not only be concerned with the relative costs of brown vs. green consumption faced by individuals. They must also address all those motivational barriers that prevent the individual agent from making the green choice, even if that choice is economically rational in a diachronic perspective (\cite{A2007E,Pirni}). In other words, there may be cases where the classical instruments of climate policy, such as subsidies and taxes, are not sufficient to induce individual agents to switch from brown to green; agents may also need to be provided with the means to overcome the ``impatience of the present", which often leads them to postpone choosing what is economically rational in the medium-term future (\cite{A2010Book}). One such means is intertemporal nudging, i.e. the temporal separation between making a green choice and actually paying for it (see, e.g., \cite{E2000CUP, M2015book}). Indeed, if the hyperbolic discounting of the future utility of a green choice is associated with the approaching cost of the choice, policymakers could allow agents to commit to the choice at a time before the time at which the cost is borne. In addition to subsidising green energy and increasing the cost of fossil energy, policymakers could encourage firms to allow consumers to commit to buying green products, as in the case of the PV systems analysed here, at a point in time prior to the actual purchase (\cite{CF2021CC}). Further research is needed on how much time should elapse between the time of commitment and the time of purchase.\\
\indent The article highlights that motivational aspects are both an obstacle that should not be overlooked, but also an opportunity for policymakers, as nudging is usually associated with little or no cost (\cite{LM2019}). This fact is quantified in Scenarios II and III in Section 4. In particular, in Scenario II, we show how, given two economic policies with equal economic benefits, the motivational/irrational aspects lead an agent to choose one over the other and thus indicate which policy would be more effective. In Scenario III, on the other hand, we quantify the effectiveness of a policy based on nudging and show that it could be as effective as a policy based on incentives or price reductions.\\
\indent The study also provides empirical insights relevant to individual climate ethics. Those who take a consequentialist view hold that the decision to consume green or brown must be made on the basis of the sum of the consequences, both negative and positive, that the choice generates in the world – rather than, for example, on the basis of the motivations for the choice, or the virtues that the individual may cultivate through the choice, etc. (see, e.g., \cite{garvey2011,jamieson2007}). The main obstacle to justifying a duty to consume green from a consequentialist perspective is the problem of causal inefficacy. The latter consists in the fact that if the individual agent goes green with a single act of consumption, e.g. buying a PV system, this will have no or imperceptible effect on the mathematics of global warming: with or without the additional PV system, global temperature projections would be the same 
 (e.g., \cite{KS2018ETMR} and \cite{F2016CC}). Individuals, therefore, when choosing what to consume, should limit themselves to choosing what is best for them. Rather than changing individual consumption patterns, climate change obligations will consist of promoting change at the political level (e.g., \cite{M2019TheMonist} and \cite{sardo2022}).\\
\indent The causal inefficacy argument is based on the assumption that the choice between green and brown consumption is mainly an economic decision made by individuals in isolation from other people’s decisions. The study shows that this assumption is not as robust as it is usually made out to be. In fact, the results help to show that individual consumer choices matter not only for the GHG emissions that the agent reduces by making the green choice, but also for the effect that the individual green choice has on the likelihood that other individuals will make similar choices (\cite{frank2021under}). The results in Section \ref{sec:numerical_experiments}, in particular Scenario IV, help to bring out this more complex picture and contribute to changing the way in which individual mitigation duties should be formulated.\\
\indent Individual behavior has social spill-over effects that go far beyond the individual abatement potential. In fact, the individual who makes the green choice contributes more to climate change mitigation, not only through the GHG emissions avoided, but also through the social signal it sends. It could even be argued that, in some cases, communicating a green choice may have a greater climate impact than actually making it. Indeed, communicating the green choice can help to overcome various barriers on the part of those who have not yet switched from fossil to green energy, e.g. barriers related to technical knowledge or economic viability, or it can even trigger social pressure mechanisms that are naturally reinforced as the number of actors who have made the green choice increases. Accordingly, individuals need to signal that they have made the green choice in as many ways as possible. There are, of course, many different ways of doing this, ranging from traditional word-of-mouth, to interaction via the internet and social networks, to simple display.\\  
\indent Furthermore, the attempt of providing an effective communication related to the green choice might inaugurate a twofold positive effect, from the point of view of the nudging theory. On the one hand, it can become a kind of self-rewarding - and therefore further motivating - action: the relevant result just achieved could motivate the single agent in pursuing other ``behavioural shifts" in favor of other ``green choice" (e.g.: a different selection in food purchase, or a more efficient different ways for management using  of water consumption in the domestic domain). On the other hand, an effective communication of that choice might ``nudge" other people, or could function as an exemplary model for other subjects that might replicate the same choice.}\\
\section{Conclusions}\label{sec:conclusions}
\textcolor{black}{The present paper aims to analyze the diffusion of residential PV systems in Italy from 2006 to 2026. Our approach involves a Markovian agent-based model that takes into account the influence of network communication and the payback period of the investment on adoption decisions. The agents in our model are \textit{bounded rational homo economicus} agents and share similar interactions with each other and the environment, as well as decision rules characterized by a small number of parameters. This simplification allows us to summarize agent behaviour using macroscopic laws and equations through a mean field approach. We found that it was relatively easy to calibrate the model by adjusting the weights of the innovation-decision process. By altering key parameters, we evaluated the projected diffusion via a scenario analysis. Our comprehensive study led to the discussion of two societal implications: firstly, cognitive biases, such as hyperbolic discounting of future utility, significantly impact green choices over time; and secondly, individual green choices influence the likelihood that others will make similar choices and reduce greenhouse gas emissions.\\ 
\indent Of course, the model is built on simplifications and assumptions and may not capture unforeseen events. However, its ability to match the actual diffusion of residential PV systems in Italy over the considered period is promising. Additionally, the proposed framework's applicability to other countries and, with minor adjustments, to other renewable energy technologies, suggests potential for future implementations with an improved set of underlying parameters.}

\newpage

\appendix

\newpage
\appendix
\section{The development of the PVs in Italy}\label{app:subsec:evolution_PVs}
\textcolor{black}{This section provides an overview of the various measures introduced by the Italian government to promote the adoption of photovoltaic (PV) systems from 2005 to the present. These measures, collectively known as ``Conto Energia" (CE), offer fixed-term contracts with favorable conditions for PV systems connected to the grid with a minimum peak power of 1kW. Local electricity providers are obligated by law to purchase the electricity generated by these PV systems. The initial CE commenced in 2005 and was structured as a net metering plan (``scambio sul posto") tailored to small PV systems. This plan aimed to encourage the direct consumption of self-generated electricity. In addition to payments for each unit of electricity produced, consumers received supplementary incentives for utilizing the self-generated energy directly. Subsequent iterations, such as CE2, were open to all PV systems but were primarily designed for larger installations with limited or no direct self-consumption of electricity. Under these later versions, the electricity produced was sold to the local energy supplier, and the CE guaranteed an additional Feed-in Tariff (FiT). It is worth noting that with each new version of the CE, the FiT decreased from 0.36 \euro/kWh in 2006 to 0.20 \euro/kWh in 2012. The implementation of CE4 in 2011 introduced financial rewards for direct energy consumption, while CE5 offered incentives based on the energy supplied to the grid along with a premium rate for self-consumption.\\
\indent Following the conclusion of the fifth CE program, the FiT and premium schemes were discontinued, and a tax credit program was introduced in 2013. This was followed by the reintroduction of a new incentive decree for photovoltaic systems (RES1) in 2019, which was explicitly reserved for systems with a capacity of more than 20 kW but not exceeding 1 MW. Subsidies were provided based on the net electricity produced and fed into the grid, with the unit incentive varying according to the plant size.\\
\indent In May 2020, the Italian government enacted the ``Revival Decree" (Decree Law 34/2020), which introduced a further increase of 110$\%$ in tax deductions. The extent of the tax deduction depended on whether the installation was associated with energy-saving measures. Moreover, the decree allowed for the free transfer of surplus energy to the grid and included provisions for subsidized tax deductions and implementing battery energy storage systems.\\
\indent Additionally, the Ministerial Decree of September 16, 2020, introduced an incentive measure, offering incentives for collective self-consumption configurations and renewable energy communities. These incentives, amounting to 100 \euro/MWh and 110 \euro/MWh, respectively, are valid for 20 years and are subject to certain limitations.}

\newpage
\section{Computation of the economic cost and of the cash flows}\label{app:investment_costs_and_cash_flows}
In this section, we describe the computation of the investment cost and the cash flows. This computation is based on \cite{PSM2015TFSS}, except for the computation of the investment costs $I_{\text{econ}}$. Indeed, in \cite{PSM2015TFSS}, the authors compute the investment costs by using a very precise formula involving the maximum peak power of the PV system, the available rooftop area for PV modules, the efficiency of the solar cells, the PV system efficiency, and the irradiation at standard conditions. Because in the present work we are trying to explain the general public’s behaviour on the GET, we have judged the previous procedure too refined, and follow the ones in \cite{KSB2022Energy}, where they provide a formula that enables the calculation of the cost of a given PV plant, on condition that the per-unit power module cost and the plant size are known. More precisely:
\begin{equation}\label{eq:Iecon}
    \begin{split}
    I_{\text{econ}}(t) =& \text{Modules Cost} [\text{Euro/Wp}]\\
                        &\times \text{Plant Size} [\text{Wp}] \times \text{Index ``Plant Cost compared to Modules Cost"}.
    \end{split}
\end{equation}
The numerical values for the previous quantities are provided in the cited reference.\\ 
\indent We now turn to the computation of the cash flow, and we start from the $R_{\text{Save}}(s, \text{CE})$. \cite{PSM2015TFSS} provide an explicit expression for $R_{\text{Save}}(s, \text{CE})$ in the case of the \text{CE} 5: 
\begin{equation*}
\begin{split}
    R_{\text{Save}}(s, \text{CE} 5)   &= E_{\text{PV}}(s) \cdot \left[\chi_{\text{DC}} \cdot p_{\text{elec},\text{buy}} \cdot (1+\tau_{\text{elec},\text{buy}})^{s-t-1} \right.\\
           &\left.+ (1-\chi_{\text{DC}}) \cdot p_{\text{elec},\text{sell}} \cdot (1+\tau_{\text{elec},\text{sell}})^{s-t-1}\right],\quad t \leq s \leq \tau.\\
\end{split}
\end{equation*}
where $E_{\text{PV}}(s)$ is the produced amount of electricity, $\chi_{\text{DC}}$ the share of direct electricity consumption and $p_{\text{elec}, \text{buy}}$ (resp. $p_{\text{elec}, \text{sell}}$) is the price of electricity bought (sold). The amount of electricity $E_{\text{PV}}$ generated by the system is a function of the level of irradiation $(E_{\text{Sun}})$, of the installed nominal maximum peak power $(P_{\text{MPP}})$, and of the predicted PV module abrasion $(\xi_{\text{Abrasion}})$:
\begin{equation*}
    E_{\text{PV}}(s) = E_{\text{Sun}} \cdot P_{\text{MPP}} \cdot (1-\xi_{\text{Abrasion}})^{s-t-1}.
\end{equation*}

Besides energy savings, an additional positive cash flow is generated by governmental support ($R_{\text{Gov}}(t, CE)$), which is based on the \text{FiT} (Feed in Tariff) given by the \text{CE}. The amount of the support is calculated as the sum of three components: a basic payment for the production of electricity ($\text{FiT}_{\text{Prod}}(CE)$), an incentive for direct PV electricity consumption ($\text{FiT}_{\text{DC}}(\text{CE})$), and, if applicable, additional bonuses ($\text{FiT}_{\text{Bon}}(\text{CE})$) that accrue
in special circumstances. The cash flows associated with
governmental support are then expressed as follows:
\begin{equation}
    R_{\text{Gov}}(s, \text{CE}) = E_{\text{PV}}(s) \cdot (\text{FiT}_{\text{Prod}}(\text{CE}) + \text{FiT}_{\text{DC}}(\text{CE}) + \text{FiT}_{\text{Bon}}(\text{CE}))
\end{equation}
\noindent For instance in the \text{CE 5} the governmental support is 200 Euro/kW and then decreases by either 15 $\%$ every six months, 5 $\%$ every six months and 25 $\%$ every six months. When there is no incompatibility with the benefits from the CEs, we add the additional benefits from the Net Metering scheme. After the end of the CEs, we consider the tax credit program, which consists of ten tax refunds, one for every year, of the size of a percentage of the initial investment, as done in \cite{ODG2015}.  We do not consider additional bonuses due to more invasive house renovations, as we suppose only a fraction of adopters could benefit from those bonuses.\\
\indent As in \cite{PSM2015TFSS}, we assume that the adoption of a PV system also entails a series of negative cash flows. Administrative fees ($R_{\text{Adm}}( CE )$) have to be paid to
the provider of the electricity grid and depend on the specific
CE considered. For example, for \text{CE 5} we have that:
\begin{equation*}
    R_{\text{Adm}}(\text{CE}) = 3 \frac{\text{Euro}}{\text{kW} \cdot \text{year}}.
\end{equation*}
Maintenance costs ($R_{\text{Main}}(t)$) must also be
considered. Upfront costs (e.g., the consultation of a PV expert/
adviser) are paid in the first year of the investment, while
maintenance costs occur yearly. Both expenditures are estimated to be a fraction of the initial investment costs (as done in \cite{PSM2015TFSS}):
\begin{equation}
    R_{\text{Main}}(s) = 
    \begin{cases}
    (\alpha_{\text{upfront}} + \alpha_{\text{Main}}) \cdot I_{\text{econ}}\quad\text{if}\quad s=t\\
    \,\alpha_{\text{Main}} \cdot I_{\text{econ}}\quad\quad\quad\quad\quad\quad\,\,\,\text{otherwise}.
    \end{cases}
\end{equation}
Finally, the cash flow includes depreciation allowance payments
of the PV system ($R_{\text{Deprec}}(s)$). The depreciation allowance
amounts to a fixed outflow taking place at the end of every year
for 20 years, at which point the remaining value of the fixed
asset at the end of its useful lifetime is zero.\\

\section{Description of the Italian \emph{Sinus-Milieus}$^{\tiny{\text{\textregistered}}}$ categories adopted in the present paper}\label{app:SinusMilieus}
The following Table \ref{tab:TableSinusMilieus} report the description of the \emph{Sinus-Milieus}$^{\tiny{\text{\textregistered}}}$ categories used in the present study. The source is Appendix A.1, Table 11, in \cite{PSM2015TFSS}.

\begin{center}
\begin{table}
\begin{tabular}{ |l|l| }
  \hline
  \textbf{\emph{Sinus-Milieus}$^{\tiny{\text{\textregistered}}}$} & \textbf{\emph{Borghesia Illuminata (enlightened middle
class)}}\\
   Characteristics      &\small{Highest lifestyle, society's elite, econ. thinking}                \\ 
   Type of household    &Couples, sometimes with children   \\
   Age                  &Older than 45 years   \\
   Education            &Highest education   \\
   Work                 &Businessmen, qualified employees and executives   \\
   Income               &Highest income   \\
   Share of population  &5.7 million inhabitants (10$\%$ of population)   \\
  \hline
    \textbf{\emph{Sinus-Milieus}$^{\tiny{\text{\textregistered}}}$} & \textbf{\emph{Progressisti Tolleranti (intellectuals)}}\\
   Characteristics      &Critical intellectuals, socially ambitious   \\ 
   Type of household    &Couples, sometimes with children   \\
   Age                  &40--60 year   \\
   Education            &High and highest education   \\
   Work                 &Freelance, executive employees   \\
   Income               &Freelance, executive employees   \\
   Share of population  &5.7 million inhabitants (10$\%$ of population)   \\
  \hline
      \textbf{\emph{Sinus-Milieus}$^{\tiny{\text{\textregistered}}}$} & \textbf{\emph{Edonisti Ribelli (experimentalists)}}\\
   Characteristics      &Modern and creative, open to
new ideas   \\ 
   Type of household    &Small families and singles   \\
   Age                  &Younger than 35 years   \\
   Education            &Higher education   \\
   Work                 &Freelancer, executive employees   \\
   Income               &Average income   \\
   Share of population  &4.1million inhabitants (7$\%$ of population)   \\
   \hline
      \textbf{\emph{Sinus-Milieus}$^{\tiny{\text{\textregistered}}}$} & \textbf{\emph{Italia Media Ambiziosa (modern mainstream)}}\\
   Characteristics      &Modern mainstream, living the social norms   \\ 
   Type of household    &Small families and singles   \\
   Age                  &All age classes  \\
   Education            &Average education  \\
   Work                 &Employees, craftsman  \\
   Income               &Average income   \\
   Share of population  &9.7 million inhabitants (17$\%$ of population)   \\
   \hline
     \textbf{\emph{Sinus-Milieus}$^{\tiny{\text{\textregistered}}}$} & \textbf{\emph{Neo Achievers (modern performers)}}\\
   Characteristics      &Performance oriented, seeking individual fulfillment   \\ 
   Type of household    &Singles, mostly male  \\
   Age                  &Younger than 35 years \\
   Education            &High education  \\
   Work                 &Freelance, specialized employees  \\
   Income               &Average to high income   \\
   Share of population  &6.4 million inhabitants (11$\%$ of population)   \\
   \hline
\end{tabular}
\caption{Detailed description of the Italian \emph{Sinus-Milieus}$^{\tiny{\text{\textregistered}}}$ categories adopted in the present paper.}
\label{tab:TableSinusMilieus}
\end{table}
\end{center}

\end{document}